\documentclass[11pt, a4paper, logo, onecolumn, address]{deepmind}

\usepackage{blindtext}
\usepackage{verbatim}
\usepackage{xcolor}
\usepackage[sorting=none]{biblatex}
\usepackage{graphicx}
\usepackage{subfig}
\usepackage{amsmath}
\usepackage{amssymb}

\title{A multi-agent reinforcement learning model of reputation and cooperation in human groups}

\author[*1]{Kevin R. McKee}
\author[*1]{Edward Hughes}
\author[1]{Tina O. Zhu}
\author[1]{Martin J. Chadwick}
\author[1]{Raphael Koster}
\author[1]{Antonio Garc{\'\i}a Casta{\~n}eda}
\author[1]{Charlie Beattie}
\author[1,2]{Thore Graepel}
\author[1,3]{Matt Botvinick}
\author[$\dagger$1]{Joel Z. Leibo}
\affil[*]{These authors contributed equally to this work}
\affil[$\dagger$]{Corresponding author}
\affil[1]{DeepMind, London, UK}
\affil[2]{Department of Computer Science, UCL, London, UK}
\affil[3]{Gatsby Computational Neuroscience Unit, UCL, London, UK}

\begin{document}

\begin{abstract}\end{abstract}

\maketitle

\addtocontents{toc}{\protect\setcounter{tocdepth}{0}}

\noindent \textbf{Collective action demands that individuals efficiently coordinate how much, where, and when to cooperate. Laboratory experiments have extensively explored the first part of this process, demonstrating that a variety of social-cognitive mechanisms influence how much individuals choose to invest in group efforts. However, experimental research has been unable to shed light on how social cognitive mechanisms contribute to the where and when of collective action.
We build and test a computational model of human behavior in Clean Up, a social dilemma task popular in multi-agent reinforcement learning research.
We show that human groups effectively cooperate in Clean Up when they can identify group members and track reputations over time, but fail to organize under conditions of anonymity. A multi-agent reinforcement learning model of reputation demonstrates the same difference in cooperation under conditions of identifiability and anonymity.
In addition, the model accurately predicts spatial and temporal patterns of group behavior: in this public goods dilemma, the intrinsic motivation for reputation catalyzes the development of a non-territorial, turn-taking strategy to coordinate collective action.}

\section{Introduction}

Efficient group coordination is a crucial underpinning of collective action. Communities that successfully undertake collective action rely on rules and conventions delineating not just \textit{how much} individuals should contribute to group efforts, but crucially \textit{where} and \textit{when} individuals should contribute \cite{janssen2010introducing}.
Human communities often rely on spatial coordination strategies, such as territoriality and privatization, to maintain irrigation and fishery systems \cite{acheson1981anthropology, wilson1994chaos} and to provide community services \cite{besley2003centralized}.
In other such social-ecological systems, temporal coordination strategies such as turn-taking and rotation schemes are central to the sustainable harvesting of natural resources \cite{berkes1986local, ostrom1990governing, trawick2001successfully}.
Spatial and temporal strategies similarly appear to coordinate group behavior in animal collectives \cite{heinsohn1995complex, kruuk1992scent, mcgowan1989sentinel, rasa1986coordinated}.
How such coordination strategies emerge is a major unanswered question for cooperation research \cite{ostrom1990governing, ostrom2005understanding}.

Laboratory research has identified a range of social cognitive mechanisms that guide individuals to act prosocially \cite{charness2002understanding, rilling2011neuroscience, sanfey2007social, van2013psychology}. However, these studies have {not mapped} the effects of such mechanisms (i.e., how much for individuals to cooperate) onto the emergence of group coordination strategies (i.e., where and when for individuals to cooperate). {The laboratory has primarily focused on} abstract economic games where the main task for participants is to choose how much to cooperate \cite{ariely2009doing, hardy2006nice, van2010cooperation, wedekind2000cooperation, wu2016reputation, yoeli2013powering}. Tasks with rich spatial and temporal dynamics have been used in {a small percentage} of prior {experimental} studies {(e.g., \cite{janssen2010lab, janssen2014experimental})}. Unfortunately, such tasks are not amenable to analysis with traditional modeling tools, impeding progress on the question of how various individual-level factors contribute to the emergence of group strategies.

Multi-agent deep reinforcement learning provides a natural approach for addressing problems with rich spatial and temporal dynamics, including strategy games like Go \cite{silver2017mastering}, Capture the Flag \cite{jaderberg2019human}, and StarCraft \cite{vinyals2019grandmaster}. Classical multi-agent reinforcement learning algorithms perform well in these competitive contexts but---due to their selfish maximization of reward---fail to resolve collective action problems \cite{leibo2017multi, perolat2017multi}. Intrinsic motivation \cite{singh2005intrinsically} offers a method to augment these algorithms with social preferences, echoing the social cognitive processes observed in humans (e.g., \cite{balliet2009social, cialdini2004social, fehr1999theory}). Algorithms with such intrinsic motivation can succeed at achieving cooperation, overcoming the difficulties posed by incentive conflict and spatiotemporal complexity \cite{eccles2019learning, foerster2018learning, hughes2018inequity, jaques2019social, mckee2020social, peysakhovich2018prosocial}.

In contrast with extensive application of reinforcement learning algorithms to model \textit{individual} human decision making \cite{botvinick2020deep, glimcher2013neuroeconomics}, multi-agent reinforcement learning has not found widespread use in scientific models of processes within human \textit{groups}.
The present work applies multi-agent deep reinforcement learning to construct a computational model of the temporal and spatial dynamics of human cooperation. In particular, we explore how an intrinsic motivation for reputation (i.e., a social cognitive mechanism) affects the emergence of cooperation and coordination strategies within groups.

\begin{figure}[t!]
    \centering
    \subfloat[(a)]{\includegraphics[height=5cm]{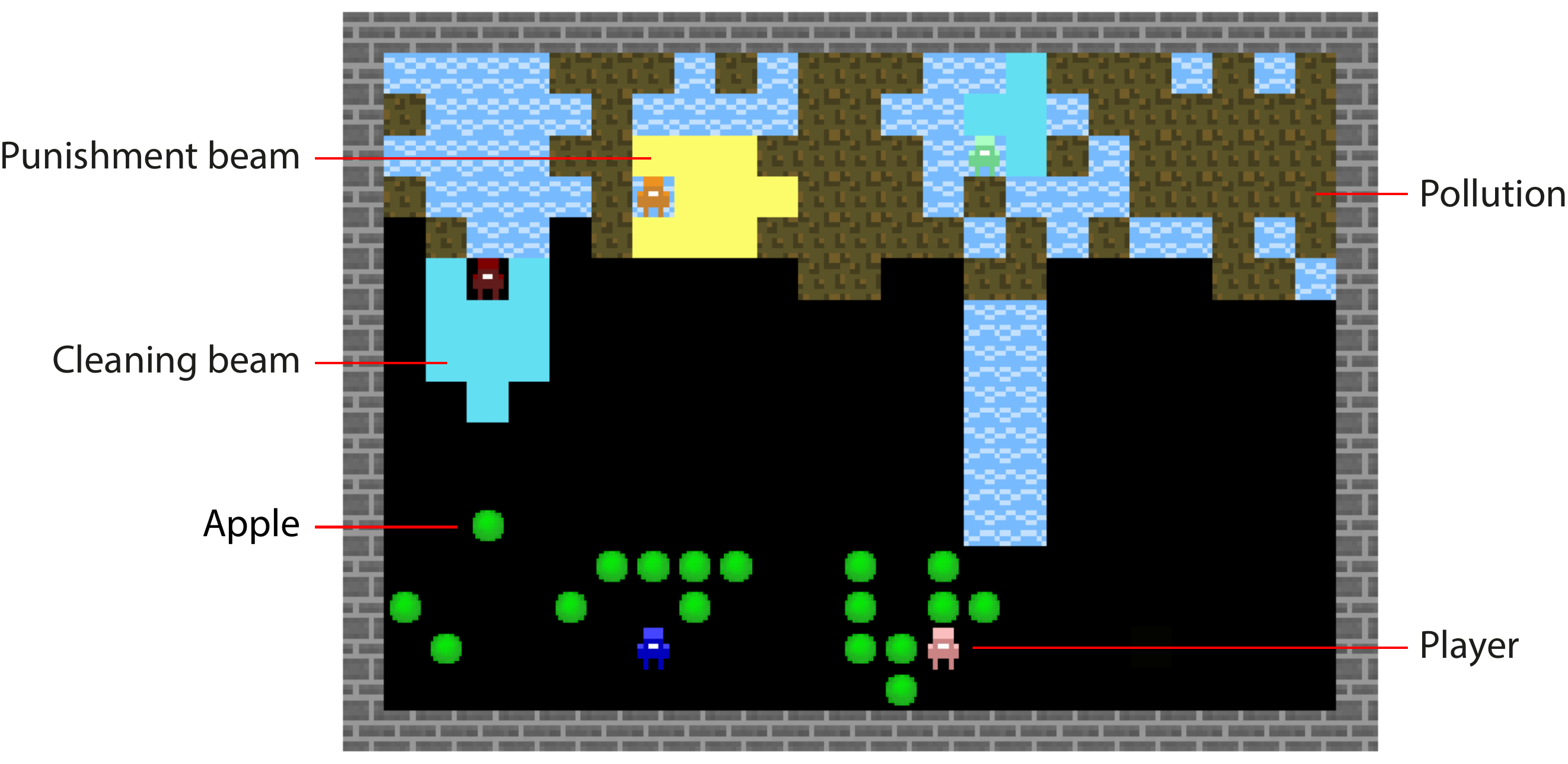} \label{fig:environment/a}}
    ~ 
    \subfloat[(b)]{\includegraphics[height=5.65cm]{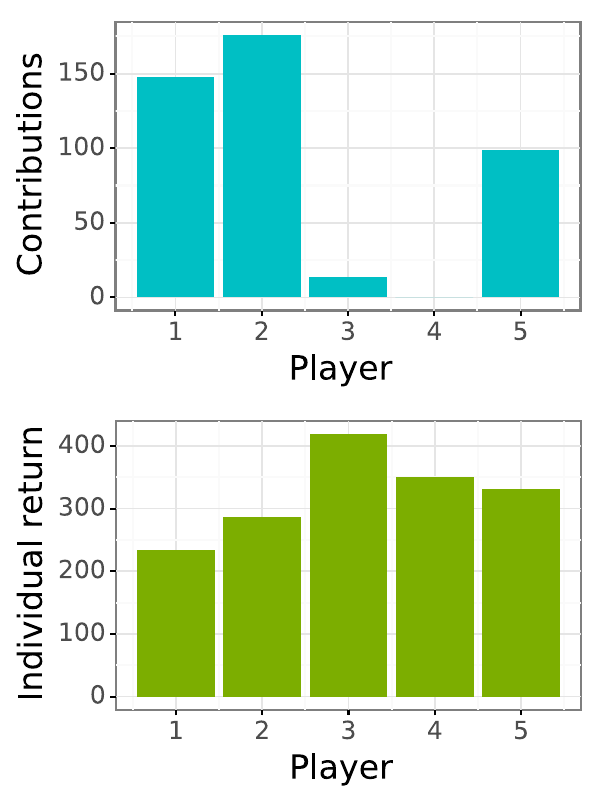}  \label{fig:environment/b}}
    \caption{Figure \ref*{fig:environment}: Clean Up is a public goods task that allows for the exploration of distributional, spatial, and temporal dynamics of collective action (i.e., \textit{how much}, \textit{where}, and \textit{when} individuals contribute to group efforts). (a) On each time step, group members ($n = 5)$ take actions in a shared environment. Individuals are rewarded for collecting apples in an orchard. Apples regrow less quickly throughout the orchard the more pollution there is in the river. Pollution accumulates stochastically in the river at a constant rate. Individuals have the ability to clean pollution and punish other group members. (b) Variations in group member behavior and rewards emerge in each episode of Clean Up (sample episode from human data). Because of the spatial separation between the river and the orchard, there is a general tradeoff between contributions and individual return.}
    \label{fig:environment}
\end{figure}

A large number of multi-agent studies have simulated group cooperation in Clean Up, a public goods task in a 2-dimensional, video game-like environment (Figure \ref{fig:environment/a}; \cite{eccles2019learning, hughes2018inequity, jaques2019social, mckee2020social, baker2020emergent, radke2022importance, leibo2021scalable, christofferson2022get, tilbury2022identity}). Clean Up draws inspiration from social dilemmas in the real world \cite{trawick2001successfully, ostrom1993coping, mollinga2003waterfront, janssen2007robustness}, incorporating greater spatial and temporal complexity than the abstracted public goods games typically studied in behavioral economics \cite{janssen2010introducing, camerer2003behavioral}. In Clean Up, group members---referred to as ``agents’’ in reinforcement learning research---are all embodied in a virtual world. They do not face a binary cooperate-defect choice, and are not tasked with selecting a single scalar number of how much to contribute.
Instead, they choose motor actions (e.g., move forward, move left, pause) based on their observation of their environment. Depending on the context, such actions may ``add up’’ to sequences that correspond with cooperation, defection, and other higher-level strategies.
Each episode of Clean Up places a small group ($n = 5$) into a two-dimensional environment, measuring $23 \times 16$ spaces and consisting of an orchard and a river. The aim of each individual group member is to collect apples from the orchard. Each group member receives one reward for every apple they collect. Apples can regrow after they are harvested. Apple growth is driven by the cleanliness of the geographically separate river. The river fills up with pollution with a constant probability over time. As the proportion of the river filled with pollution increases, the growth rate of apples monotonically decreases. If they physically leave the orchard and move to the river, group members can clear away bits of the pollution, thus contributing to a public good of cleanliness. Group members are also able to punish their peers. Unlike prior tasks, group members must put active effort into collecting the benefits of the public good by harvesting apples from the orchard. Individuals must balance the costs and benefits of time spent harvesting apples and contributing to the public good
(Figure \ref{fig:environment/b}). Critically, preliminary computational studies have indicated that a diverse range of coordination strategies suffice to resolve the collective action problem in Clean Up, drawing to varying degrees on spatial \cite{eccles2019learning, hughes2018inequity} and temporal solutions \cite{mckee2020social}. Clean Up thus provides an ideal setting to explore whether social cognitive mechanisms can steer groups toward greater cooperation and particular coordination strategies.

We constructed a computational model of group behavior in public goods dilemmas with advantage actor-critic \cite{mnih2016asynchronous, sutton1998introduction}, a commonly used deep reinforcement learning algorithm (see \textit{Materials and Methods}). Like other reinforcement learning agents, the advantage actor-critic algorithm places a strong emphasis on the importance of reward for guiding behavior and decision-making---echoing the importance placed upon reward in behavioral economics \cite{parkes2015economic}.
The algorithm's behavior is influenced by both extrinsic reward (externally determined incentives; e.g., the payoff dictated by the rules of a game or task) and intrinsic reward (internally generated incentives \cite{singh2005intrinsically}; cf., satisfaction generated by social cognitive mechanisms in the brain).
In this experiment, we investigated the intrinsic motivation for reputation \cite{berridge2009dissecting, izuma2008processing} and its influence on group coordination strategies. The core of our computational model is the overall reward signal $r$, comprising the extrinsic reward $r_e$ and the intrinsic reward $r_i$:
\begin{gather}
   r = r_e + r_i \, , \\
   r_i = - \alpha \cdot \textrm{max}(\bar{c} - c_{\textrm{self}}, 0) - \beta \cdot \textrm{max}(c_{\textrm{self}} - \bar{c}, 0) \, .
   \label{intrinsic_reward_function}
\end{gather}

\noindent Here $c_{\textrm{self}}$ is one's own contribution level (i.e., the amount of pollution cleaned from the river), $\bar{c}$ is an estimated or observed average of the group's contribution levels, and $\alpha$ and $\beta$ are scalar parameters. The $\alpha$ and $\beta$ parameters control how much $r_i$ is affected by one's score falling behind and rising above the group average score, respectively. {This function echoes the utility functions used in prior analytic \cite{fehr1999theory}, agent-based \cite{janssen2012evolution}, and reinforcement learning \cite{hughes2018inequity} models comparing individual behavior to group norms.} {For the experiments here, we parameterize the model with $\alpha \sim \mathcal{U}(2.4, 3.0)$ and $\beta \sim \mathcal{U}(0.16, 0.20)$.} The intrinsic motivation for reputation reflects primarily an aversion to having a lower reputation than one's peers, which potentially decreases one's desirability as a partner or associate \cite{hardy2006nice, barclay2006partner}. Secondarily, the intrinsic motivation reflects an aversion to peers taking advantage of one's efforts, since exploitability diminishes one's fitness relative to others \cite{rockenbach2006efficient}.

\section{Results}

We simulated rounds of Clean Up with our computational model and compared its behavior against the behavior of groups of human participants in Clean Up. In both the model and the behavioral experiment with humans, we test the effects of the motivation for reputation by comparing two conditions: (1) an \textit{identifiable} condition, in which group members are individually distinguishable and thus contribution behavior is common knowledge,\footnote{See a human group playing an identifiable round here: \url{https://youtu.be/ohQrN46n9sQ}.} and (2) an \textit{anonymous} condition, in which group members are largely indistinguishable and cannot perfectly monitor contribution behavior\footnote{See a human group playing an anonymous round here: \url{https://youtu.be/AqCKDibiE9Q}.}.
Identifiability allows the intrinsic motivation for reputation to act, whereas anonymity mitigates its influence on decision making.

After training $N = 120$ artificial agents with multi-agent reinforcement learning for our model, we simulated rounds of Clean Up and compared the agents' behavior against the behavior of groups of human participants ($N = 120$) in Clean Up. For further details on the correspondence between the computational model and behavioral experiment design, see \textit{Materials and Methods}.

\begin{figure}
    \centering
    \subfloat[(a)]{\includegraphics[height=4.85cm]{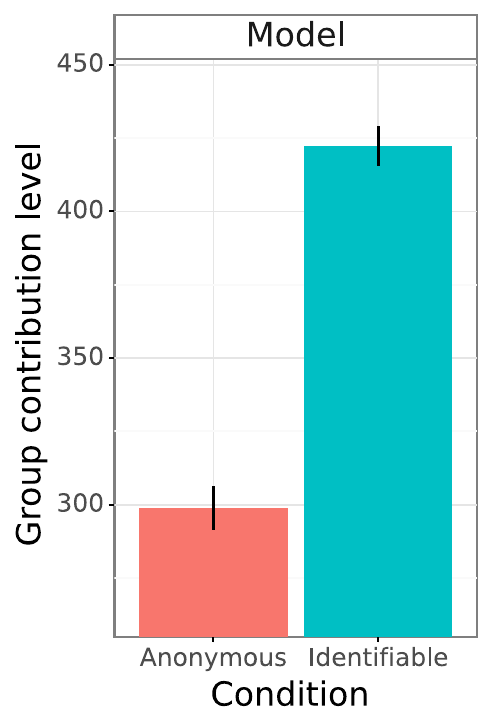}
    \includegraphics[height=4.85cm]{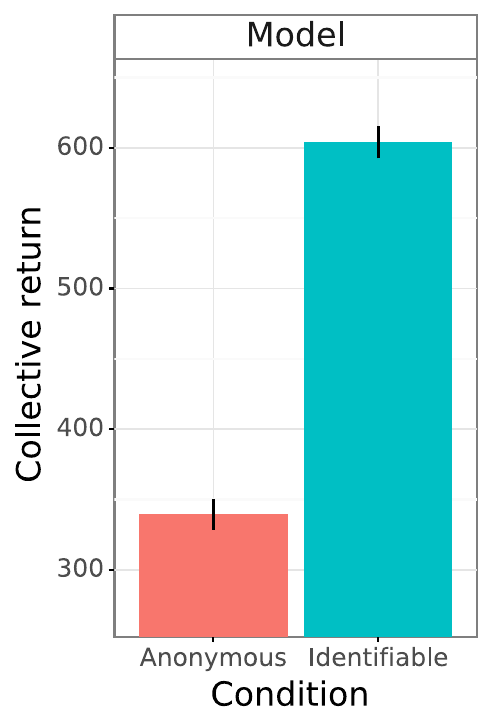} \label{fig:model_results/a}} \\
    \subfloat[(b)]{\includegraphics[height=4.5cm]{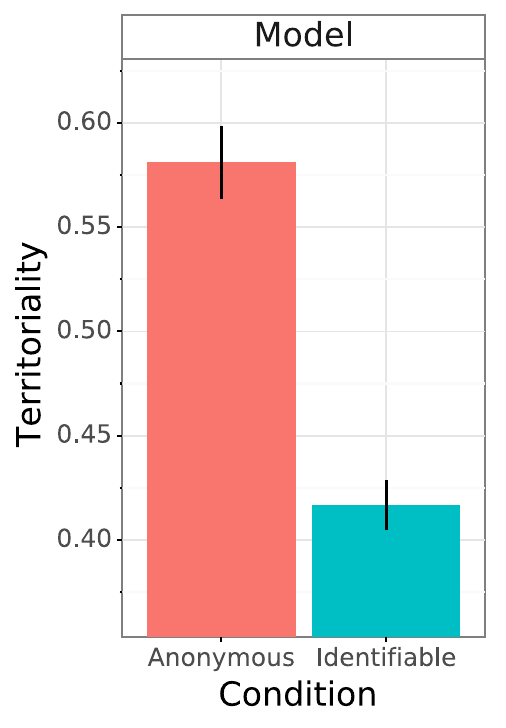}  \hspace{1em} \includegraphics[height=4.5cm]{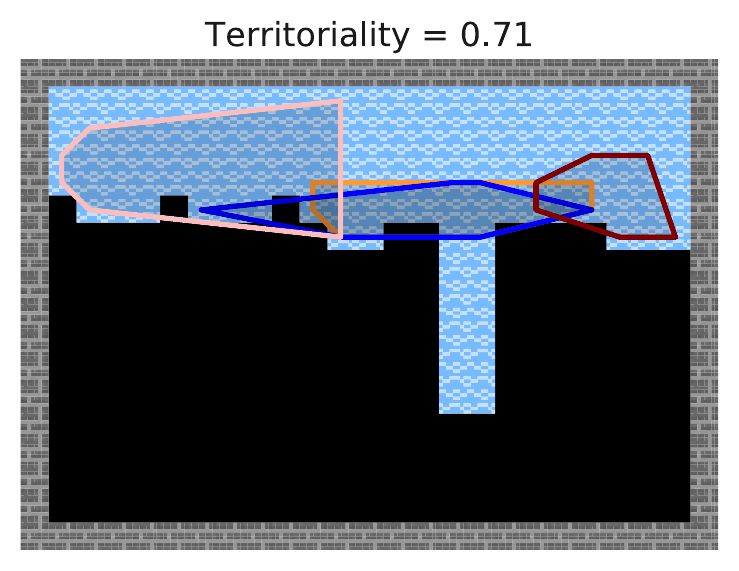} \:
    \includegraphics[height=4.5cm]{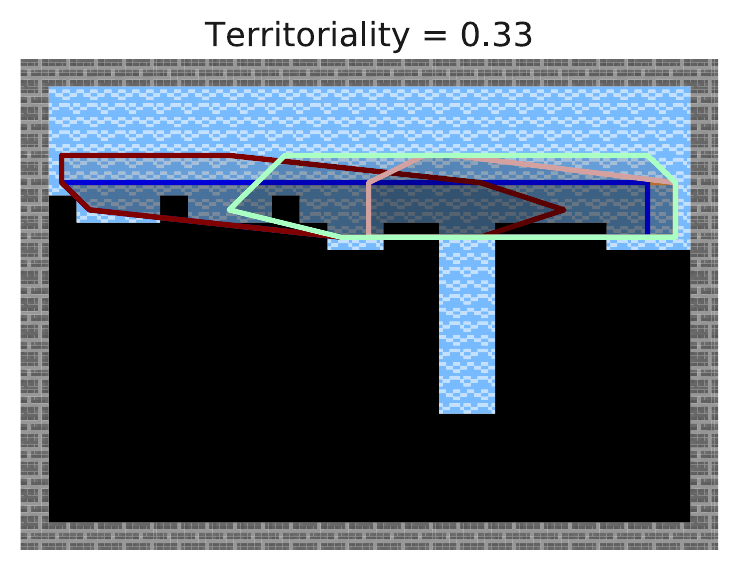} \label{fig:model_results/b}} \\
    \subfloat[(c)]{\includegraphics[height=4.5cm]{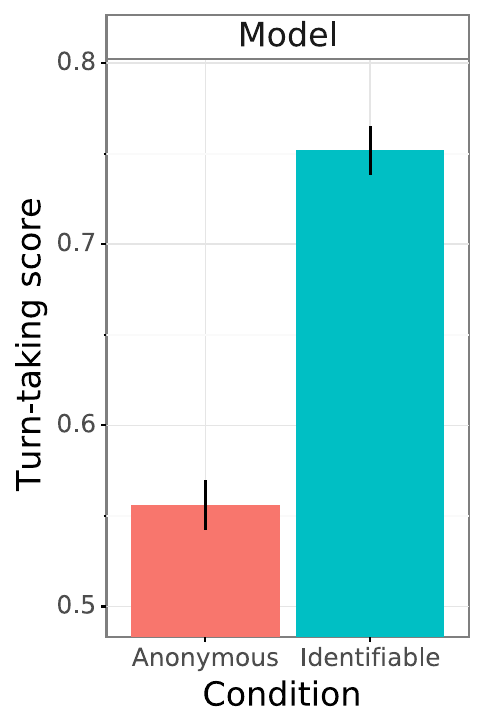} \hspace{1em}
    \includegraphics[height=4.25cm]{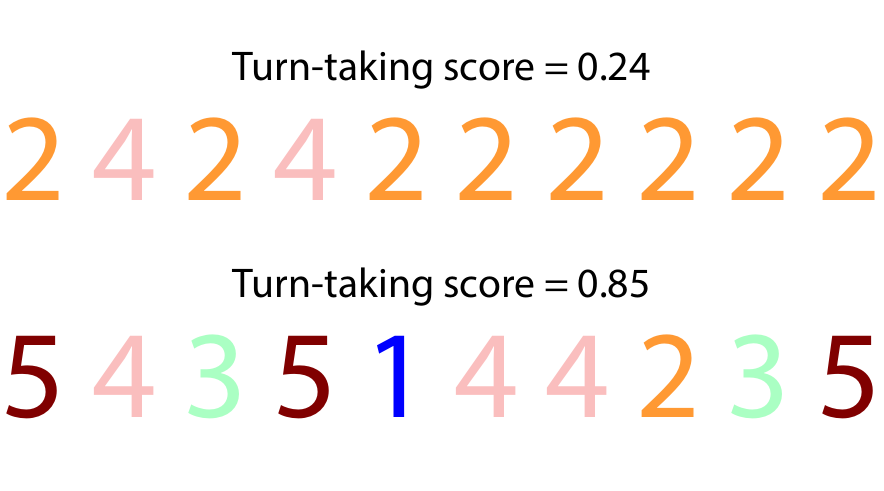} \hspace{1em}
    \includegraphics[height=4.5cm]{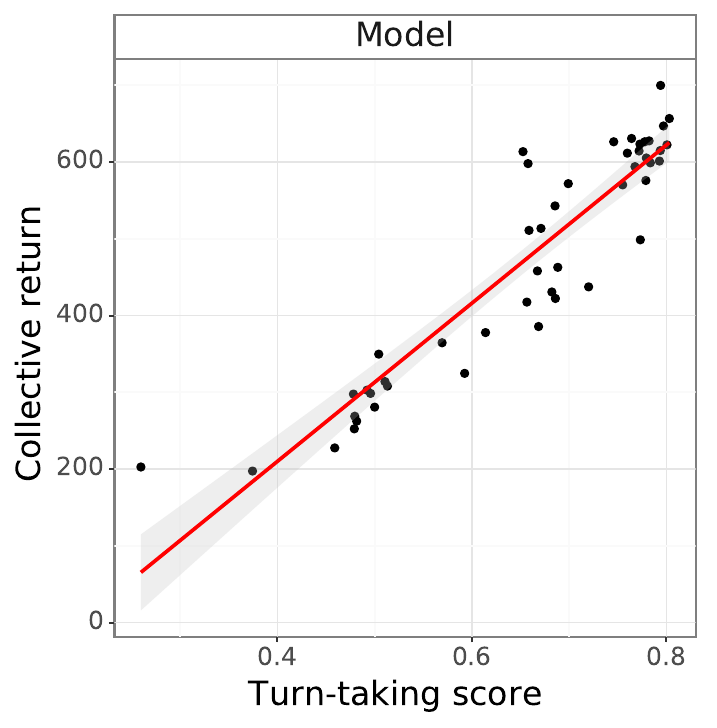} \label{fig:model_results/c}}
    \caption{Figure \ref*{fig:model_results}: The intrinsic motivation for reputation substantially alters the behavior of deep reinforcement learning agents in Clean Up. Here we report the $F$ ratio, degrees of freedom, and $p$-value for each ANOVA. Errors bars reflect 95\% confidence intervals. (a) In the identifiable condition, when the motivation for reputation exerted a strong influence on behavior, there were significant increases in both average contribution level, $F(1,\:311) = 1090.7$ ($p < 0.0001$), and collective return, $F(1,\:311) = 2030.3$ ($p < 0.0001$). (b) Under conditions of identifiability, groups were significantly less territorial in the river, $F(1,\:311) = 432.0$ ($p < 0.0001$). To the right are example patterns of river ``territories'' from the model, showing a group that exhibited high territoriality ($\textrm{territoriality} = 0.71$) and a group that exhibited lower territoriality ($\textrm{territoriality} = 0.33$). (c) In contrast, groups were significantly \textit{more} reliant on turn taking in the identifiable condition, $F(1,\:311) = 758.3$ ($p < 0.0001$). Example patterns of turn taking from the model are included here, listing the identities of the group members that took the first ten turns of the episode (i.e., entering into the river to clean). The examples include a group that exhibited low turn taking ($\textrm{turn taking} = 0.24$) and a group that exhibited higher turn taking ($\textrm{turn taking} = 0.85$). These turn-taking scores were significantly correlated with {group performance, $\beta = 1030.3$, $p < 0.0001$.}}
    \label{fig:model_results}
\end{figure}

We begin by validating the effects of the intrinsic motivation for reputation on group outcomes in our computational model. As expected, identifiability produced a significant increase in group contribution levels, $p < 0.0001$ (repeated-measures ANOVA, Figure \ref{fig:model_results/a}). This increase in contribution levels led to significantly higher collective returns, $p < 0.0001$ (repeated-measures ANOVA, Figure \ref{fig:model_results/a}). When motivated to cultivate a good reputation, group members in the model increase their cooperativeness, resulting in higher payoffs for the entire group.

Having confirmed that our computational model successfully captures the effect of reputation on \textit{how much} to cooperate, we next leverage the model to make novel predictions about the emergent spatial and temporal dynamics of group behavior. We focus on two behavioral questions informed by research on social-ecological systems. Specifically, we ask whether groups of humans sustain cooperation by allocating responsibility for public goods maintenance through spatial territories (``territoriality''; \cite{janssen2008turfs}) or through temporal rotation schemes (``turn taking''; \cite{ostrom1994rules}). Because prior reputation studies have focused on abstract economic games, they offer no predictions for these questions.

In the model, the intrinsic motivation for reputation appears not to catalyze a territorial approach to maintaining the public good. We observe significantly more spatial overlap between group member territories in the identifiable condition than in the anonymous condition, $p < 0.0001$ (repeated-measures ANOVA, Figure \ref{fig:model_results/b}). Example patterns of high and low territoriality from the computational model are presented in \ref{fig:model_results/b}.

Instead, groups in the model appear to coordinate their efforts with a temporal rotation scheme. After measuring the extent to which group members' take ``turns'' entering and cleaning the river, we observe significantly greater turn taking under identifiable conditions---when the intrinsic motivation for reputation influences behavior---than under anonymous conditions, $p < 0.0001$ (repeated-measures ANOVA, Figure \ref{fig:model_results/c}). Example patterns of low and high turn taking from the computational model are presented in Figure \ref{fig:model_results/c}. Further, results from the model confirm that this turn-taking strategy is associated with higher group performance. Across episodes, the more a group relied on a turn-taking rotation scheme, the higher {the collective return it received}, $p < 0.0001$ (linear regression, Figure \ref{fig:model_results/c}).

Our computational model suggests that the intrinsic motivation for reputation should not only increase a group's average contribution level in Clean Up, but also generate a turn-taking, non-territorial strategy for coordinating group efforts.
We next evaluate the effectiveness of the model by exploring data collected from a behavioral experiment with human participants.

\begin{figure}
    \centering
    \subfloat[(a)]{\includegraphics[height=4.85cm]{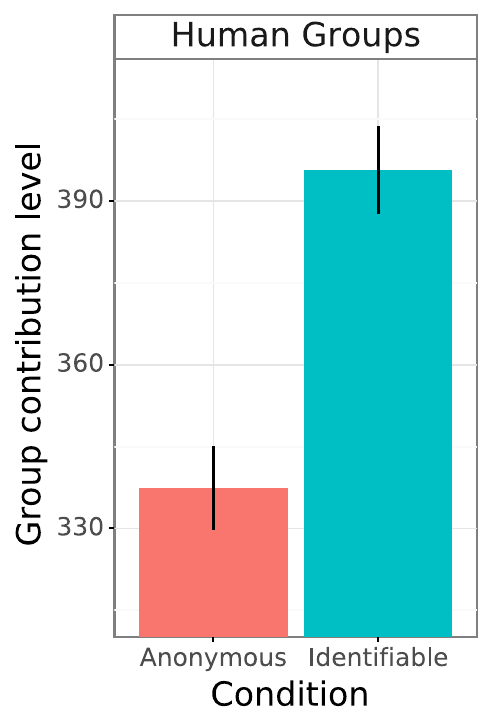}
    \includegraphics[height=4.85cm]{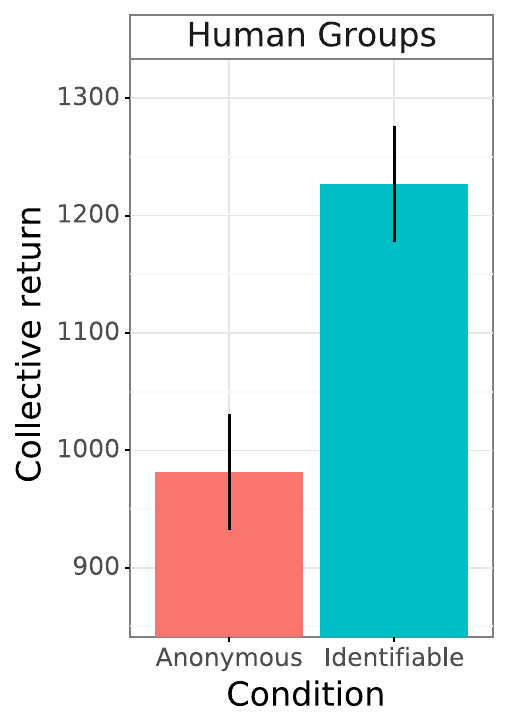} \label{fig:human_results/a}} \\
    \subfloat[(b)]{\includegraphics[height=4.5cm]{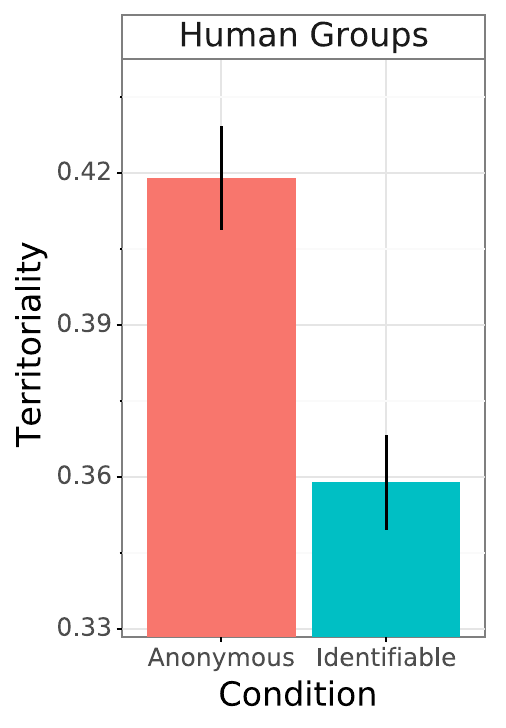} \hspace{1em}
    \includegraphics[height=4.5cm]{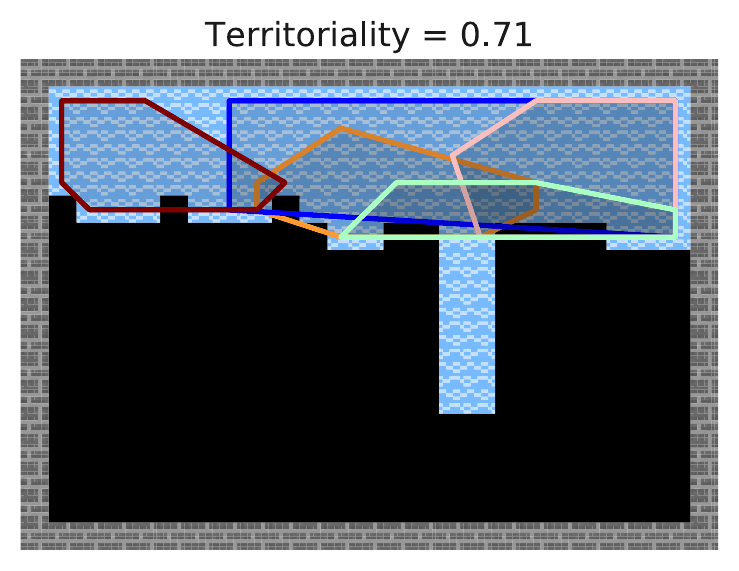} \:
    \includegraphics[height=4.5cm]{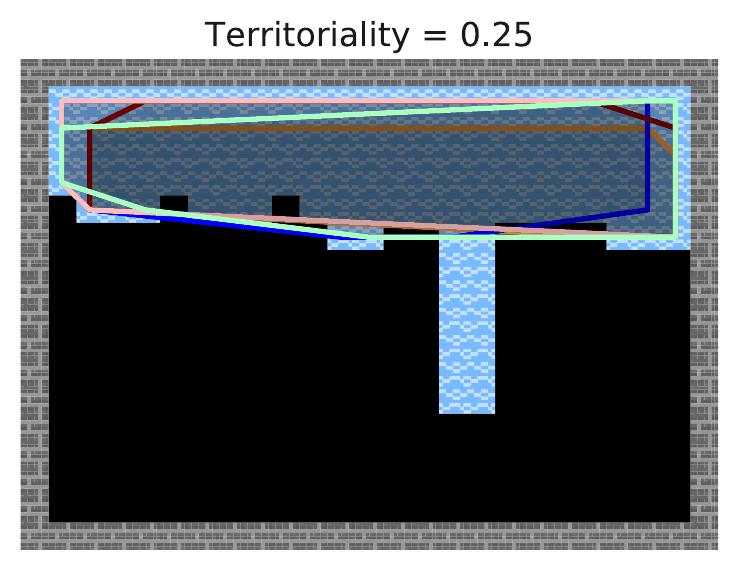} \label{fig:human_results/b}} \\
    \subfloat[(c)]{\includegraphics[height=4.5cm]{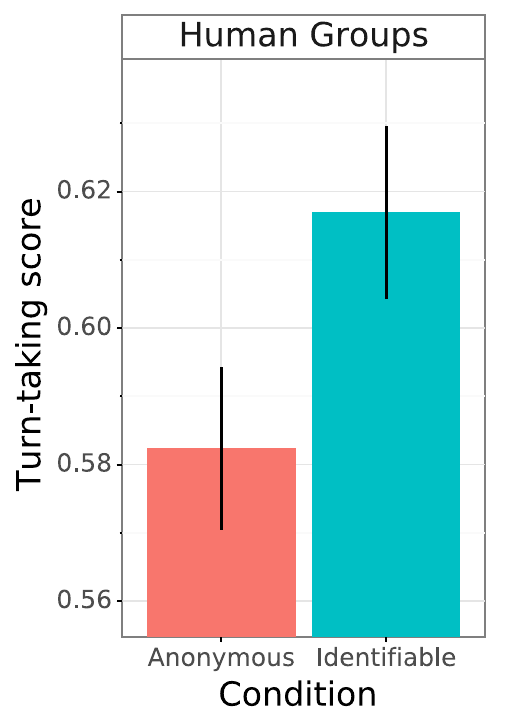} \hspace{1em}
    \includegraphics[height=4.25cm]{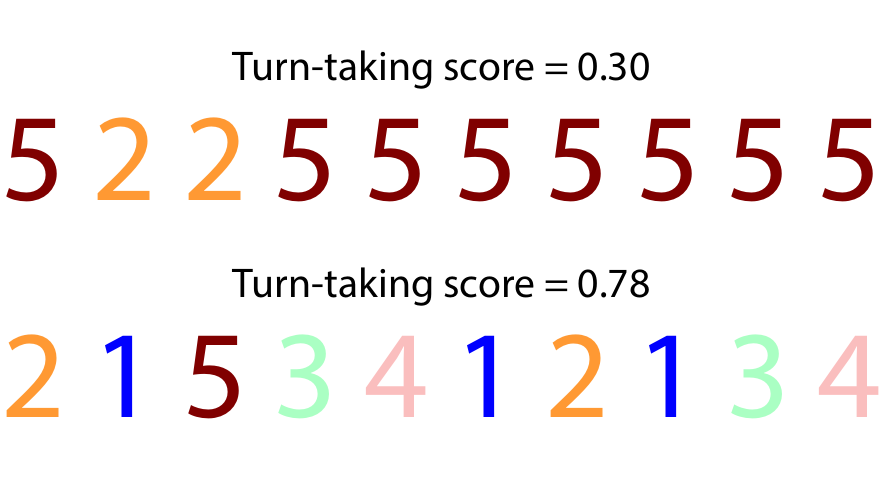} \hspace{1em}
    \includegraphics[height=4.5cm]{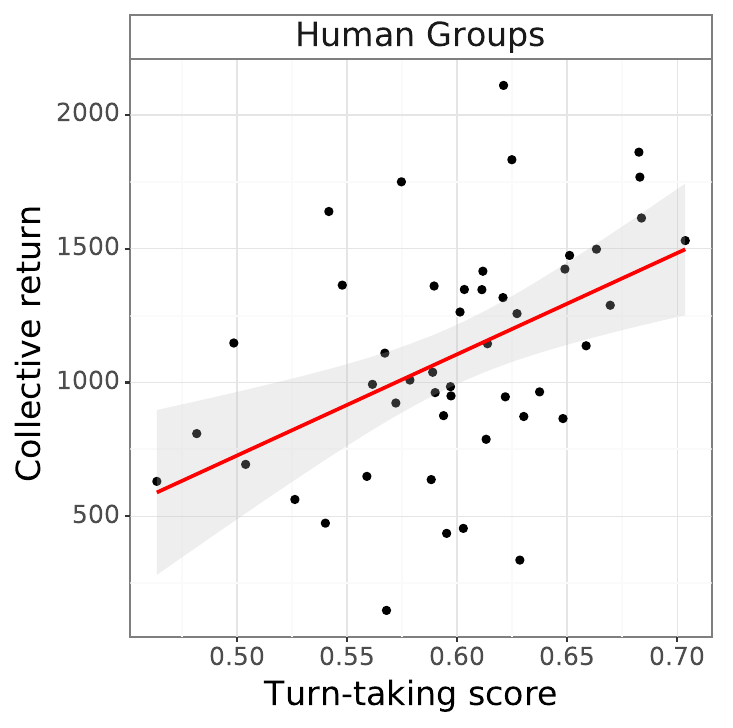} \label{fig:human_results/c}}
    \caption{Figure \ref*{fig:human_results}: The intrinsic motivation for reputation substantially alters the behavior of human groups in Clean Up, matching the predictions made by our computational model. Errors bars reflect 95\% confidence intervals. (a) In the identifiable condition, when the motivation for reputation exerted a strong influence on behavior, there were significant increases in both average contribution level, $F(1,\:310) = 199.4$ ($p < 0.0001$), and collective return, $F(1,\:310) = 89.4$ ($p < 0.0001$). (b) Under conditions of identifiability, groups were significantly less territorial in the river, $F(1,\:310) = 138.4$ ($p < 0.0001$). To the right are example patterns of river ``territories'' among human groups, showing a group that exhibited high territoriality ($\textrm{territoriality} = 0.71$) and a group that exhibited lower territoriality ($\textrm{territoriality} = 0.25$). (c) In contrast, groups were significantly \textit{more} reliant on turn taking in the identifiable condition, $F(1,\:311) = 29.4$ ($p < 0.0001$). Example patterns of turn taking from the model are included here, listing the identities of the group members that took the first ten turns of the episode (i.e., entering into the river to clean). The examples include a group that exhibited low turn taking ($\textrm{turn taking} = 0.30$) and a group that exhibited higher turn taking ($\textrm{turn taking} = 0.78$). These turn-taking scores were significantly correlated with {group performance, $\beta = 3784.6$, $p = 0.0010$.}}
    \label{fig:human_results}
\end{figure}

As expected, conditions of identifiability lead human groups to substantially increase their contribution levels. Group contribution level is significantly higher in the identifiable condition than in the anonymous condition, $p < 0.0001$ (repeated-measures ANOVA, Figure \ref{fig:human_results/a}). Similarly echoing findings from prior experiments, collective return for human groups increased significantly in the identifiable condition, $p < 0.0001$ (repeated-measures ANOVA, Figure \ref{fig:human_results/a}).

The crucial test {for the model is whether it predicts the presence of spatial or temporal strategies for human groups---the where and when of cooperation}. Consequently, we next analyze spatial and temporal patterns of group behavior in our human data.

As in the model, reputational motivation does not lead human groups to rely on spatial coordination in Clean Up. Human groups exhibited significantly less territoriality in the identifiable condition than in the anonymous condition, $p < 0.0001$ (repeated-measures ANOVA, Figure \ref{fig:human_results/b}). Example patterns of low and high territoriality from the human groups are provided in Figure \ref{fig:human_results/b}.

In contrast, participants were substantially more reliant on rotation schemes to organize collective action. There was significantly greater turn taking in the identifiable condition than in the anonymous condition of the human behavioral experiment, $p < 0.0001$ (repeated-measures ANOVA, Figure \ref{fig:human_results/c}). Example patterns of low and high turn taking from the human groups are provided in Figure \ref{fig:human_results/c}. As in the model, rotating responsibility for contributions was associated with {improved group performance. The more a group relied on a turn-taking strategy, the higher the collective return it tended to achieve}, $p = 0.0010$ (linear regression, Figure \ref{fig:human_results/c}).

\section{Discussion}

The ability to resolve collective action problems is a hallmark of human sociality. Humans do not encounter these dilemmas in a vacuum: group approaches to cooperation and defection must be coordinated over space and time. Through our evolutionary history, we have developed powerful psychological mechanisms to navigate this complex socio-physical landscape \cite{henrich2017secret, henrich2007humans, herrmann2007humans}.

In this work, we bring together behavioral research and multi-agent reinforcement learning to model the emergence of coordination strategies in public goods dilemmas. The model we propose springs from two broad lines of influence: first, behavioral research demonstrating the effects of social cognition on individual prosociality \cite{hardy2006nice, fehr1999theory, camerer2003behavioral}, and second, studies showing the capabilities of multi-agent deep reinforcement learning algorithms \cite{jaderberg2019human, vinyals2019grandmaster, baker2019emergent}. Rather than analyzing decisions in abstracted economic games, our research explores behavior in a temporally and spatially extended task. Here, cooperation is a composite strategy aggregating decisions about where and when to act, rather than a simple choice of how much to contribute. We endow the reinforcement learning agents in our model with an intrinsic motivation \cite{singh2005intrinsically} for reputation \cite{van2010cooperation, wedekind2000cooperation, nowak1998evolution, milinski2002reputation}. We use this computational model to make detailed predictions about how this intrinsic motivation influences the coordination of group behavior.

Notably, previous multi-agent studies have observed the emergence of diverse temporal and spatial strategies in Clean Up, depending on the intrinsic motivation incorporated into each group member \cite{eccles2019learning, jaques2019social, mckee2020social}. Our results suggest that the performance of multi-agent deep reinforcement learning agents can be fruitfully compared with behavior recorded from human groups on the same task.

Our model captures the substantial improvements in outcomes produced when individuals can track reputations, in accord with previous studies. Individuals increase their contributions when motivated to achieve a good reputation, resulting in significant increases in group return. The model further generates new insight into how social cognitive processes shape group strategies for cooperation. Reputation tracking catalyzes a non-territorial turn taking-like strategy, resulting in consistent maintenance of the public good.

The new approach introduced here has implications beyond the study of reputation. This framework draws our understanding of multi-agent reinforcement learning closer to our understanding of human cognition and behavior. {It expands the toolkit available to investigate and examine mechanisms of group cooperation \cite{dafoe2020open}.}
How does the temporal and spatial structure of human interaction affect our ability to solve collective action problems \cite{miller1992collective}? What intrinsic motivations can support the formation and maintenance of institutions \cite{north1991institutions}? Answers to these questions can help us scaffold collective action and strengthen cooperation in communities of humans and artificially intelligent agents.

\section{Materials and Methods}

{In addition to this overview, precise details of the computational model and human behavioral experiment are provided in the supplementary information.}

Clean Up is a public goods task in a two-dimensional, video game-like environment, measuring 23 by 16 spaces (Figure \ref{fig:environment/a}; see \cite{hughes2018inequity}). In the Clean Up task, the aim is to collect apples from an orchard. Each group member gains one point for every apple he or she collects. Apples can regrow after they are harvested; apple regrowth is driven by the cleanliness of a geographically separate river. The river fills up with pollution with a constant probability over time. As the proportion of the river filled with pollution increases, the regrowth rate of apples monotonically decreases. For sufficiently high pollution levels, no apples will regrow.

Group members have a water beam tool which allows them to clean pollution from the river. The public good in Clean Up is the regrowth rate of the orchard. Group members contribute to the public good by cleaning the river. Group members also possess a ``ticketing'' tool, providing a mechanism for costly punishment \cite{fehr2002altruistic, henrich2006costly}. The ticketing tool allows them to lower the scores of other group members. It costs four points to ticket another group member; the group member receiving the ticket loses 40 points.

We train the agents in the model using independent multi-agent reinforcement learning based on policy gradients. Concretely, each agent individually comprises a deep neural network, with no parameter sharing between different agents. The inputs to the neural network are the pixels representing the agent's local view of the environment and temporally smoothed data on its own contributions and the contributions of its peers. The outputs from the network are a policy (a probability distribution over the next action to take in the environment) and a value function (an estimate of the agent's discounted future return under the policy). The network architecture consists of a convolutional neural network with $3 \times 3$ kernel, stride $1$ and $32$ output channels, a two-layer multi-layer perceptron with $64$ hidden units in each layer, a long short-term memory (LSTM) \cite{hochreiter1997long} of hidden size $128$, and linear layers for the policy logits and value function. 

In independent multi-agent reinforcement learning, each agent $i$ learns a policy $\pi^i$ intended to maximize its value from some initial state $s_0$ under the joint policy of all agents. The value for agent $i$ is defined to be $V_{\mathbf \pi}(s_0) = \mathbb{E}_{\mathbf{\pi}} \left( \sum_{t=0}^\infty \gamma^t r_t^i \right)$, where $\gamma$ is a discount factor and $r_t^i$ is a random variable representing the reward at time $t$ given the actions sampled from the stochastic policies and the stochastic transition function of the environment. We use $\gamma = 0.99$ for our experiments. Each policy $i$ is parameterized by $\theta^i$, the weights of the neural network described previously.

Policy gradient methods perform gradient ascent on the parameters of the neural network to maximize the value. The policy gradient theorem \cite{sutton2000policy, williams1992simple} provides a method for updating the policy parameters based on a sample trajectory, a set of $(s_t^i, a_t^i, r_t^i)$ tuples obtained by the agents interacting with the environment. The parameters are updated by $\Delta \theta^i = \delta \gamma^t G^i_t \nabla_{\theta^i} \log \pi^i(a^i_t|s^i_t, \theta^i)$, where $G^i_t$ is the realized return from time $t$ on the given trajectory and $\delta$ is a learning rate. To reduce variance, we use the learned value function as a baseline, replacing $G^i_t$ by $G^i_t - V^i(s_t)$ in the previous equation. This algorithm is known as advantage actor-critic. We use the RMSProp optimizer \cite{tieleman2012lecture} to compute the gradient, using learning rate $0.000321$, epsilon $10^{-5}$, momentum $0$, and decay $0.99$. To encourage exploration, we use an entropy regularizer (as in \cite{mnih2016asynchronous}) with entropy cost $0.00154$.

In the training stage of the reinforcement learning experiment, we used a distributed framework \cite{espeholt2018impala, mckee2021quantifying} to train a population of $120$ reinforcement learning agents for each condition. The parameters of the agents were stored on $120$ learner processes, each responsible for carrying out the policy gradient update for one agent. To generate experience for the agents, $2000$ parallel arenas were created. For each episode in each arena, $5$ agents were randomly sampled from the population and their parameters synchronized from their respective learners. At the end of each episode, trajectories for agents were forwarded to the respective learners. Each learner aggregated trajectories in batches of $10$ and processed these to update the parameters for the associated agent, unrolling the LSTM for $100$ steps to train the recurrent network. We augmented the advantage actor-critic algorithm with VTrace \cite{espeholt2018impala} to correct for off-policy trajectories. Each agent was trained using 100 million $(s_t^i, a_t^i, r_t^i)$ tuples.

In the evaluation stage of the reinforcement learning experiment, we partitioned each population into $24$ groups of five agents at random. We assessed the performance and behavior of each group in seven episodes of Clean Up. Groups were assigned to the identifiable or anonymous condition based on the condition they experienced during training.

We recruited 120 participants for the behavioral experiment. Participants were first individually instructed on the action controls and the environmental dynamics in Clean Up through a series of tutorial levels (for exact details, see supplementary information). Subsequently, participants were sorted into groups of $n = 5$ and progressed through {14} episodes of Clean Up: {seven episodes in each condition}. We used a counterbalanced, within-participant design, with half of the groups completing the identifiable task first and the anonymous task second, and the other half completing the anonymous task first and the identifiable task second. {Like the agents in the computational model, participants observe and act based on a local view of the Clean Up environment.} After finishing both conditions, participants completed post-task questionnaires. At the end of the experiment, participants were paid according to their performance in the task. Each point accrued was worth $\frac{1}{2}$ pence. {A detailed description of the experimental protocol (including instructions and comprehension checks) can be found in supplementary information.}

More information on the Clean Up environment, including exact parameters and additional analyses verifying its social dilemma structure, can be found in supplementary information. Details of the analyses presented in the main text can similarly be found in supplementary information.

\subsection{Code availability}

The Clean Up task is available for research use through the DeepMind Lab2D platform \cite{beattie2020deepmind}.

\section{Acknowledgements}
We thank Lucy Campbell-Gillingham, Dorothy Chou, Julia Cohen, Tom Eccles, Richard Everett, Stephen Gaffney, Ian Gemp, Demis Hassabis, Koray Kavukcuoglu, Zeb Kurth-Nelson, Vicky Langston, Brian McWilliams, Matthew Phillips, Oliver Smith, Tayfun Terzi, Gregory Thornton, and Sasha Vezhnevets for feedback and support.

\subsection{Author contributions}
J.Z.L. conceived of the overall research direction; K.R.M. managed the project; A.G.C. and C.B. designed and coded the Clean Up task; K.R.M., E.H., T.O.Z., M.J.C., R.K., and J.Z.L. designed the human behavioral experiment; K.R.M., E.H., and T.O.Z. implemented the human behavioral research protocol; K.R.M. and T.O.Z. collected data for the human behavioral experiment; K.R.M., E.H., and J.Z.L. designed and coded the reinforcement learning agents and ran the reinforcement learning experiments; K.R.M., E.H., and T.O.Z. analyzed the data, with M.J.C., R.K., T.G., and M.B. providing substantial assistance; K.R.M., E.H., and J.Z.L. wrote the manuscript, with substantial assistance from T.O.Z., M.J.C., R.K., T.G., and M.B.

\subsection{Funding}

This research was funded by DeepMind.

\subsection{Competing interests}

The authors report no competing interests.

\subsection{Correspondence and materials}

Correspondence should be addressed to J.Z.L.

\printbibliography

\clearpage

\setcounter{section}{0}
\addtocontents{toc}{\protect\setcounter{tocdepth}{2}}

\setcounter{equation}{0}
\renewcommand{\theequation}{S\arabic{equation}}

\setcounter{figure}{0}
\renewcommand{\thefigure}{S\arabic{figure}}

\setcounter{table}{0}
\renewcommand{\thetable}{S\arabic{table}}

\clearpage
\part*{Supplementary Info}

\tableofcontents

\section{Design of Computational Model}

We built our computational model using advantage actor-critic~\cite{mnih2016asynchronous}, a deep reinforcement learning algorithm. Within the algorithm, we formalize the overall reward signal $r$ as a combination of the intrinsic, social reward $r_i$ and extrinsic, environmental reward $r_e$:

\begin{gather}
   r = r_e + r_i \, ,\\
   \label{eqn:reward_function}
   r_i = - \alpha \cdot \textrm{max}(\bar{c} - c_{\textrm{self}}, 0) - \beta \cdot \textrm{max}(c_{\textrm{self}} - \bar{c}, 0) \, ,
\end{gather}

\noindent where $c_{\textrm{self}}$ is one's own contribution level (i.e., the amount of pollution cleaned from the river), $\bar{c}$ is an estimated or observed average of the group's contribution levels, and $\alpha$ and $\beta$ are scalar parameters.

\begin{figure}[ht]
    \centering
    \includegraphics[width=7.5cm]{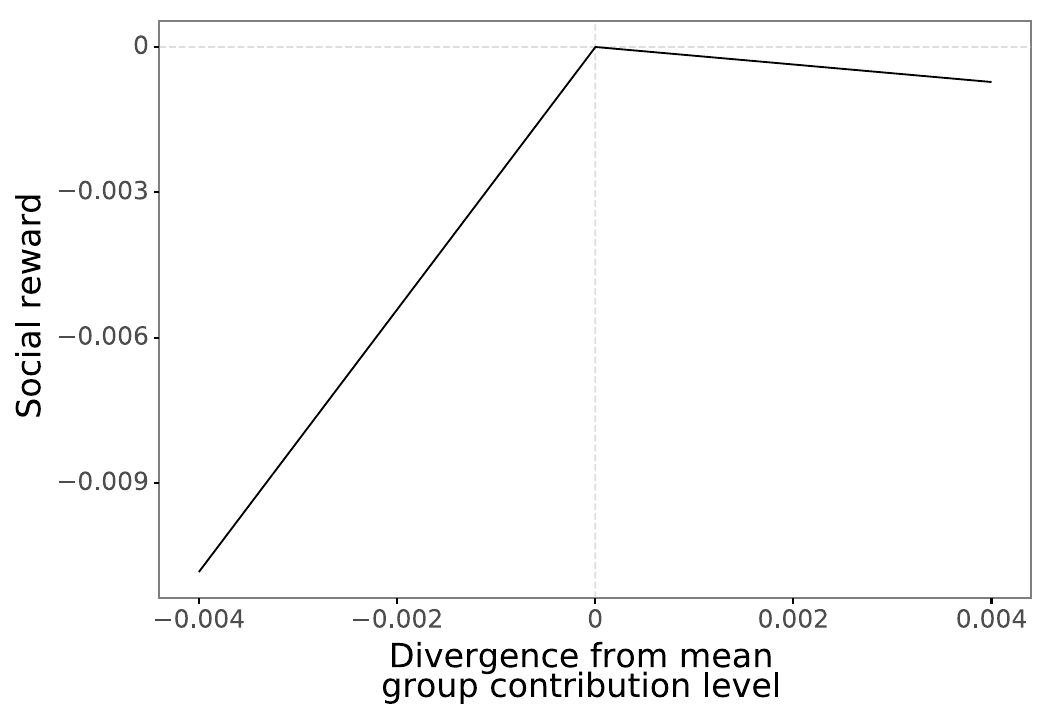}
    \caption{Figure \ref*{fig:agent_pseudoreward}: Intrinsic motivation for each reinforcement learning agent varies as a function of $c_{\textrm{self}} - \bar{c}$. Here we show the empirical mean effect for the population of agents in the model, with $\alpha \approx 2.71$ and $\beta \approx 0.18$.}
    \label{fig:agent_pseudoreward}
\end{figure}

The intrinsic motivation function for the reputation agent was parameterized with $\alpha \sim \mathcal{U}(2.4, 3.0)$ and $\beta \sim \mathcal{U}(0.16, 0.20)$ (see Figure \ref{fig:agent_pseudoreward}). It therefore primarily represents an aversion to falling behind the group mean \cite{hardy2006nice}, as well as secondarily reflecting an aversion to being cheated by others \cite{rockenbach2006efficient}.

\begin{figure}[ht]
    \centering
    \includegraphics[width=7.5cm]{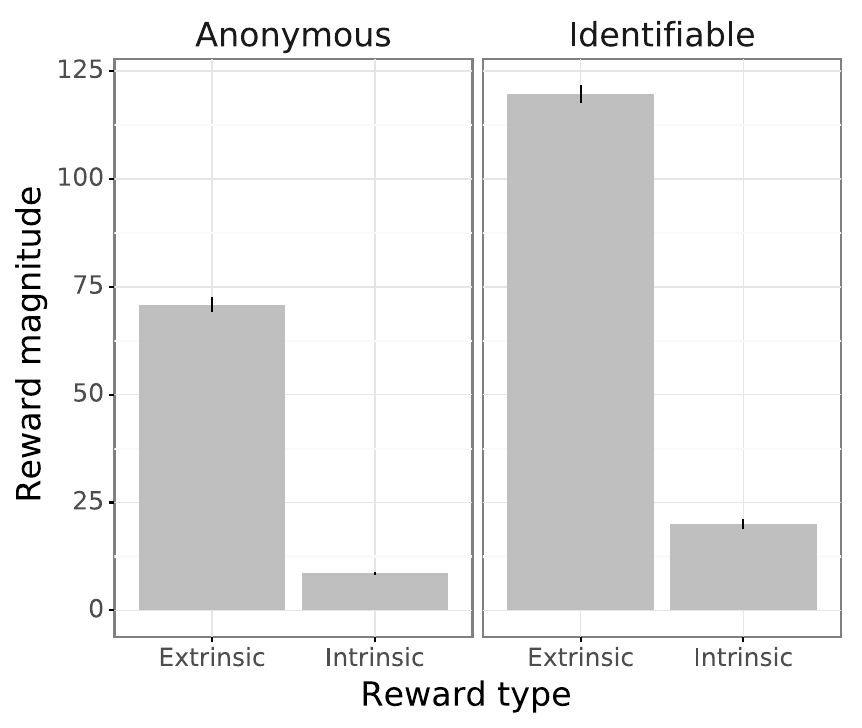}
    \caption{Figure \ref*{fig:agent_reward_comparison}: Agents in the model do not live off intrinsic reward. A paired \textit{t}-test indicates that agents received significantly less intrinsic reward than extrinsic reward per episode across all episodes of the anonymous condition and all episodes of the identifiable condition. Errors bar reflect 95\% confidence intervals.}
    \label{fig:agent_reward_comparison}
\end{figure}

A paired \textit{t}-test indicates that agents received significantly less intrinsic reward than extrinsic reward across all episodes of the anonymous condition, $t(839) = 67.8$, $p < 0.0001$ and all episodes of the identifiable condition, $t(839) = 103.6$, $p < 0.0001$ (Figure \ref{fig:agent_reward_comparison}).

We vary the intrinsic motivation parameters to better capture individual variability (cf., individual differences research; \cite{cronbach1957two}) and because population heterogeneity has important effects in multi-agent reinforcement learning \cite{mckee2020social, mckee2021quantifying}.

Let $q^t_j$ be the instantaneous index of agent contribution:
\begin{equation}
   q^t_j =     \left\{
\begin{array}{ll}
      1 & \text{agent $j$ contributed on timestep $t$} \, , \\
      0 & \text{otherwise} \, , \\
\end{array} 
\right. 
\end{equation}

\noindent and let $T$ be the number of timesteps in an episode ($T=1000$ for the reinforcement learning experiment). For agents $j = 1, \dots 5$, and the sequence of instantaneous agent contributions $\{q^t_j:t = t1 \dots t_{T}\}$, we update the temporally smoothed inputs ($c_{\textrm{self}}$ and $\bar{c}$) in Equation \ref{eqn:reward_function} as follows:

\begin{equation}
    c_j^t (c_j^{t-1}, q_j^t) = \lambda c^{t-1}_j + q^t_j\, ,
\end{equation}

\noindent where we choose smoothing factor $\lambda = 0.97$ and set $c_j^0 = 0$.

To simulate the identifiable condition for the reinforcement learning agents, the environment provides contribution information for each group member as an input to the agent. To simulate the anonymous condition, the environment provides contribution information with a reduced visibility range. Contribution information about other agents is circumscribed by a visibility range $R = 9$, calculated using the $\ell^\infty$ norm. Behavior of agents falling outside of this range is not included in updates. This follows the limited visual attention that participants marshal to track behavior in the anonymous condition (see also \cite{cohen2016bandwidth}). The provision of this altered contribution information reflects the ability of humans to imperfectly track behavior of individual group members on short timescales, even in the anonymous condition.

Overall, our computational model incorporates five assumptions:

\begin{enumerate}
    \item Behavior is motivated by the combination of extrinsic reward (externally determined incentives; e.g., the payoff dictated by the rules of a game or task) and intrinsic reward (internally generated incentives; e.g., satisfaction generated by social cognitive mechanisms in the brain \cite{singh2005intrinsically, berridge2009dissecting, izuma2008processing}).
    \item Reputation is encoded as an intrinsic reward \cite{phan2010reputation}.
    \item The intrinsic reward for reputation reflects primarily an aversion to having a lower reputation than one's peers \cite{hardy2006nice, barclay2006partner}, and secondarily an aversion to peers taking advantage of one's efforts \cite{rockenbach2006efficient}.
    \item It is common knowledge among group members which choices or actions affect reputations, and in which direction \cite{milinski2002reputation, sherif1936psychology}.
    \item Human cooperation plays out over multiple timescales, and the intrinsic motivation for reputation can operate on the timescale of minutes \cite{izuma2008processing, phan2010reputation, saxe2008love}.
\end{enumerate}

\section{Experimental Design}

Clean Up is a partially observable Markov game \cite{littman1994markov}. A small set of parameters control the Clean Up environment. Environmental dynamics are defined by two functions:

\begin{equation}
   \textrm{Pr}_{\textrm{apple}}^t = \textrm{Pr}_{\textrm{apple}} \cdot \frac{H_{\textrm{depletion}} - F_{\textrm{polluted}}^t}{H_{\textrm{depletion}} - H_{\textrm{abundance}}} \, ,\label{eqn:apple_production_function}
\end{equation}

\begin{equation}
   \textrm{Pr}_{\textrm{pollution}}^t = \textrm{Pr}_{\textrm{pollution}} \cdot (F_{\textrm{polluted}}^t < H_{\textrm{depletion}}) \, . \label{pollution_production_function}
\end{equation}

Equations \ref{eqn:apple_production_function} and \ref{pollution_production_function} describe the probabilistic production functions for apple regrowth and pollution accumulation, respectively. In Equation \ref{eqn:apple_production_function}, $\textrm{Pr}_{\textrm{apple}}^t$ represents the probability of apple regrowth at time $t$, $\textrm{Pr}_{\textrm{apple}}$ reflects the underlying probability of apple regrowth when the river is sufficiently clean, and $F_{\textrm{polluted}}^t$ represents the fraction of the river that is filled with pollution at time $t$. $H_{\textrm{depletion}}$ reflects the proportion of the river filled with pollution above which apples can no longer regrow, and $H_{\textrm{abundance}}$ reflects the proportion of the river filled with pollution below which apples regrow with maximum probability. Equation \ref{pollution_production_function} describes the Bernoulli process that generates additional pollution in the river. $\textrm{Pr}_{\textrm{pollution}}^t$ reflects the probability that a new unit of pollution accrues in the river at time $t$, while $\textrm{Pr}_{\textrm{pollution}}$ represents the underlying probability that pollution accumulates if the river is not saturated with pollution.

We ran Clean Up with {the parameter values listed in Sections \ref{sec:agent_env_params} and \ref{sec:human_env_params}}. For the reinforcement learning experiment, we drew from canonical research using this task \cite{hughes2018inequity, mckee2020social} and largely carried over the established parameters. {Because of general differences between human and agent reaction times (cf. \cite{vinyals2019grandmaster}), this approach had to be adapted to develop a human behavioral research protocol instantiating a parallel social dilemma} (see Section \ref{sec:social_dilemma_analysis}: Social Dilemma Analysis).

\subsection{Computational Model}
\label{sec:agent_env_params}

See \cite{hughes2018inequity, mckee2020social} for exact setup. The experiment was parameterized with episode length $T = 1000$ steps, cost of giving a ticket $-1$ and penalty for receiving a ticket $-50$, and the following environmental parameters:

\begin{itemize}
    \item $\textrm{Pr}_{\textrm{apple}} = 0.03\, $.
    \item $\textrm{Pr}_{\textrm{pollution}} = 0.5\, $.
    \item $H_{\textrm{abundance}} = 0.0\, $.
    \item $H_{\textrm{depletion}} = 0.32\, $.
\end{itemize}

\begin{figure}[ht]
    \centering
    \includegraphics[width=7.5cm]{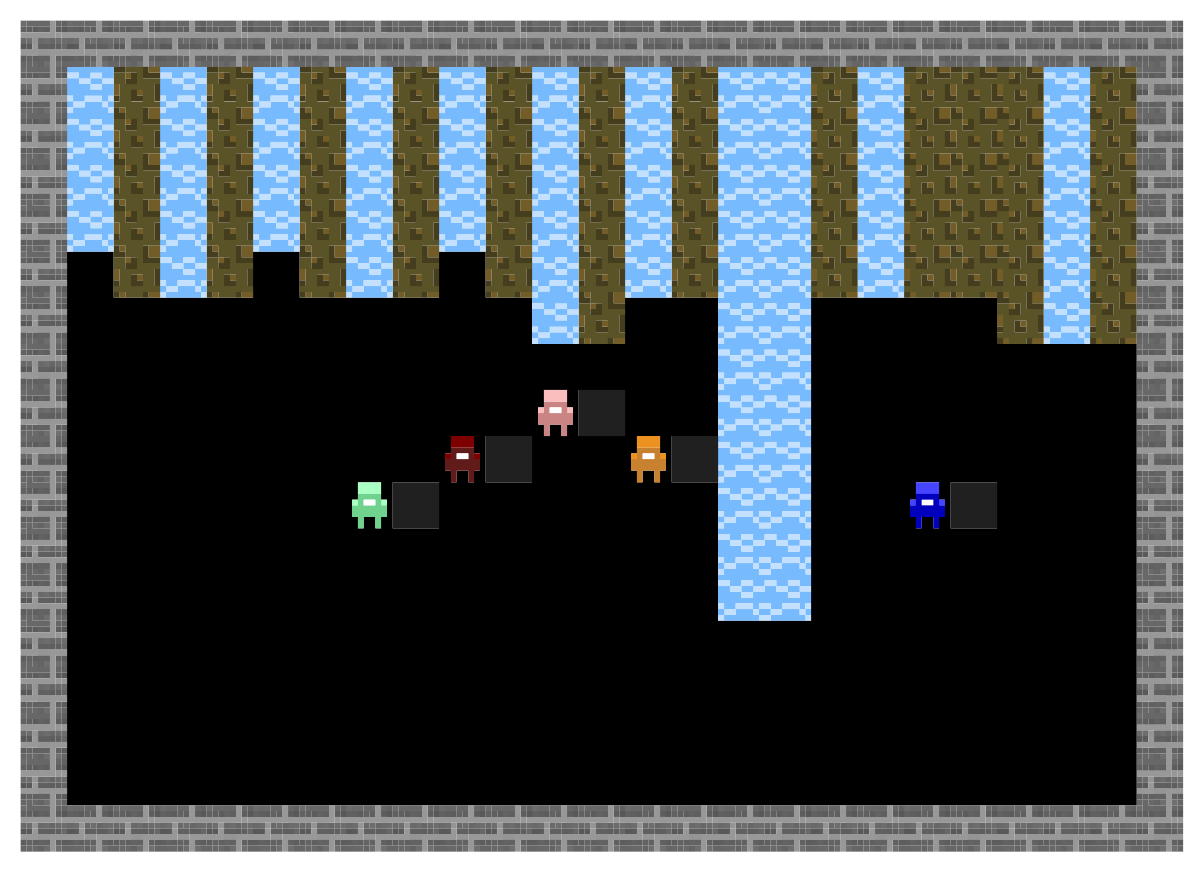}
    \caption{Figure \ref*{fig:agent_training_map}: Example initial environmental conditions for each episode of agent training. Group members were randomly initialized across various positions in the middle of the environment.}
    \label{fig:agent_training_map}
\end{figure}

Following previous studies \cite{hughes2018inequity, mckee2020social}, each episode of reinforcement learning agent training began with river pollution at saturation and an empty orchard (Figure \ref{fig:agent_training_map}). During agent evaluation, each episode began with zero river pollution and a full orchard, matching the task conditions as experienced by human participants.

\subsection{Human Behavioral Experiment}
\label{sec:human_env_params}

Consistent with the design of video games with similar step rates, we implemented an input buffer to accept actions at most once every 100 ms. {In order to instantiate an analogous social dilemma for participants}, we selected the following set of environmental parameters for the human behavioral experiment:

\begin{itemize}
    \item $\textrm{Pr}_{\textrm{apple}} = 0.067\, $.
    \item $\textrm{Pr}_{\textrm{pollution}} = 0.6\, $.
    \item $H_{\textrm{abundance}} = 0.3\, $.
    \item $H_{\textrm{depletion}} = 0.6\, $.
\end{itemize}

\begin{figure}[p]
    \centering
    \subfloat[(a)]{\includegraphics[width=10cm]{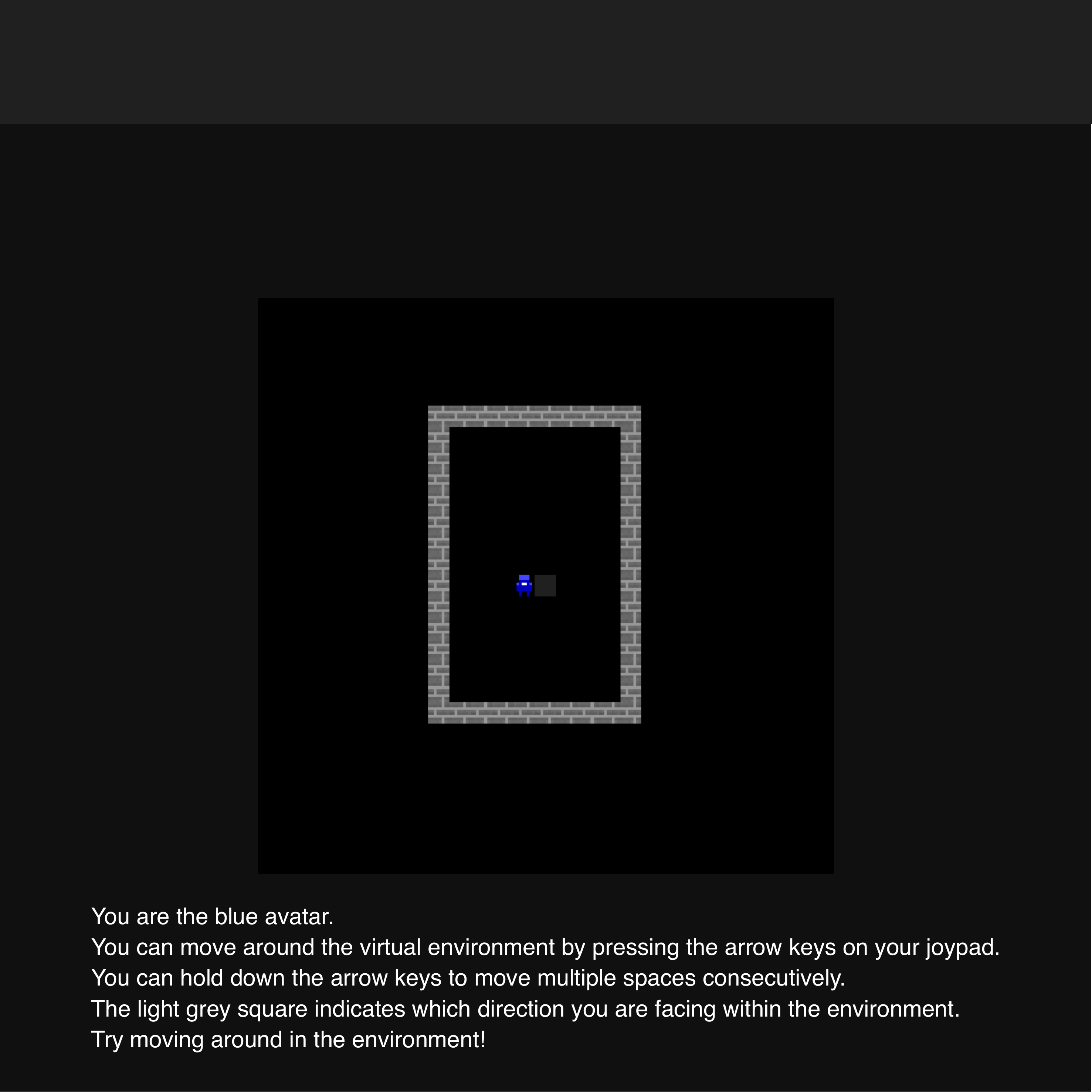}} \\
    \subfloat[(b)]{\includegraphics[width=10cm]{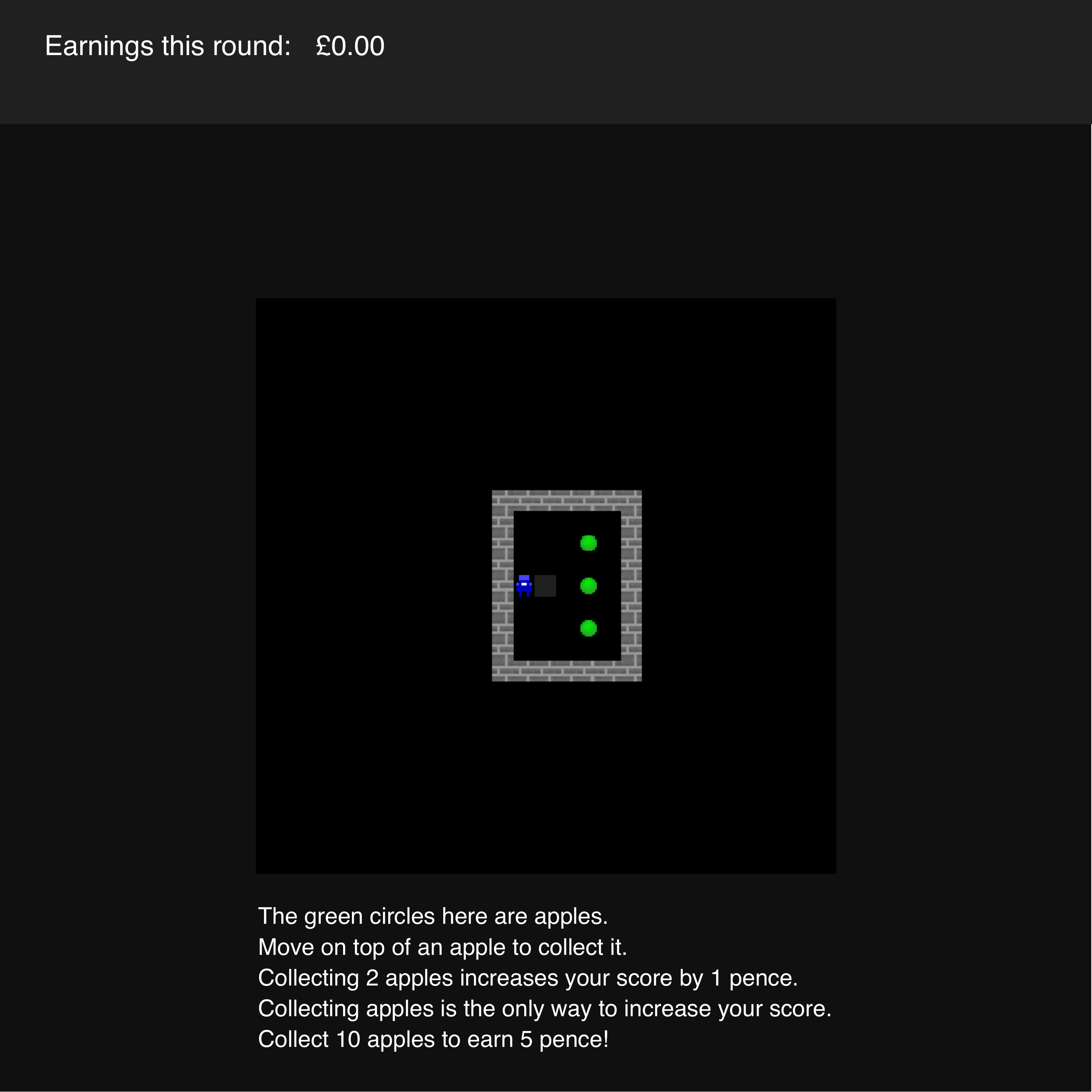}}
    \caption{Figure \ref*{fig:tutorials_a_b}a-b: Participants completed a number of tutorials to help them learn the controls for the task and the environmental dynamics of Clean Up.}
    \label{fig:tutorials_a_b}
\end{figure}

\begin{figure}[p]
    \centering
    \subfloat[(a)]{\includegraphics[width=10cm]{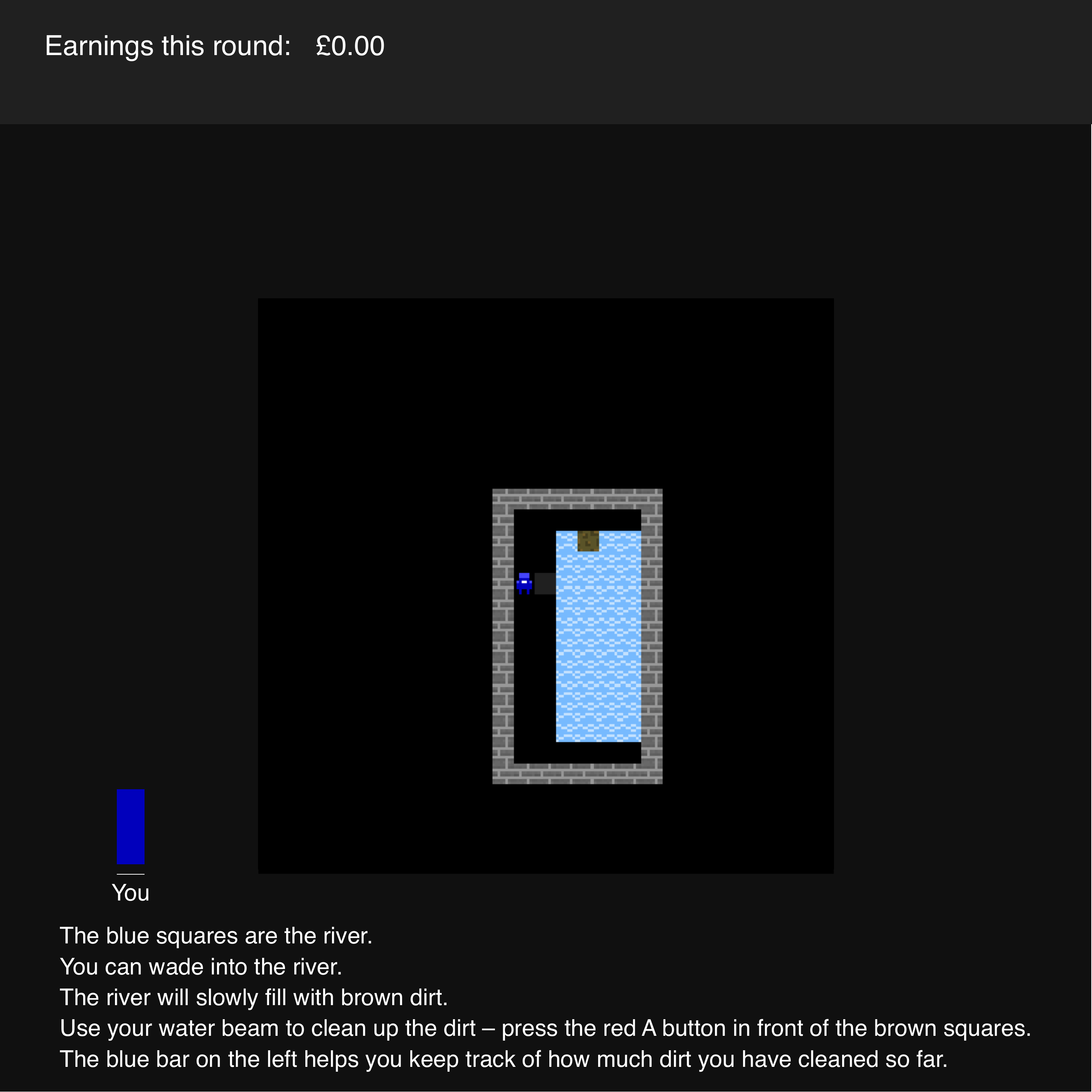}} \\
    \subfloat[(b)]{\includegraphics[width=10cm]{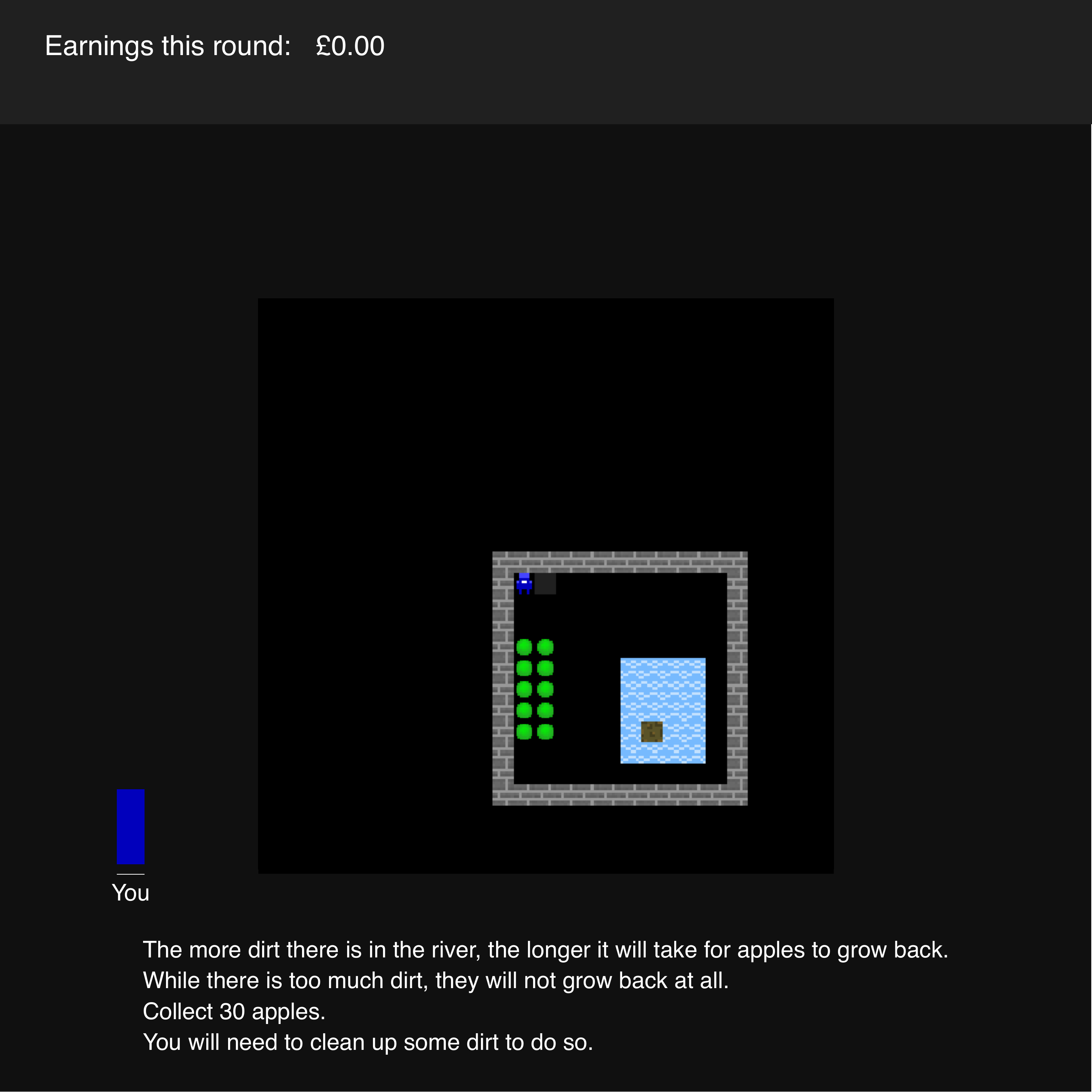}}
    \caption{Figure \ref*{fig:tutorials_c_d}a-b: Participants completed a number of tutorials to help them learn the controls for the task and the environmental dynamics of Clean Up.}
    \label{fig:tutorials_c_d}
\end{figure}

\begin{figure}[p]
    \centering
    \subfloat[(a)]{\includegraphics[width=10cm]{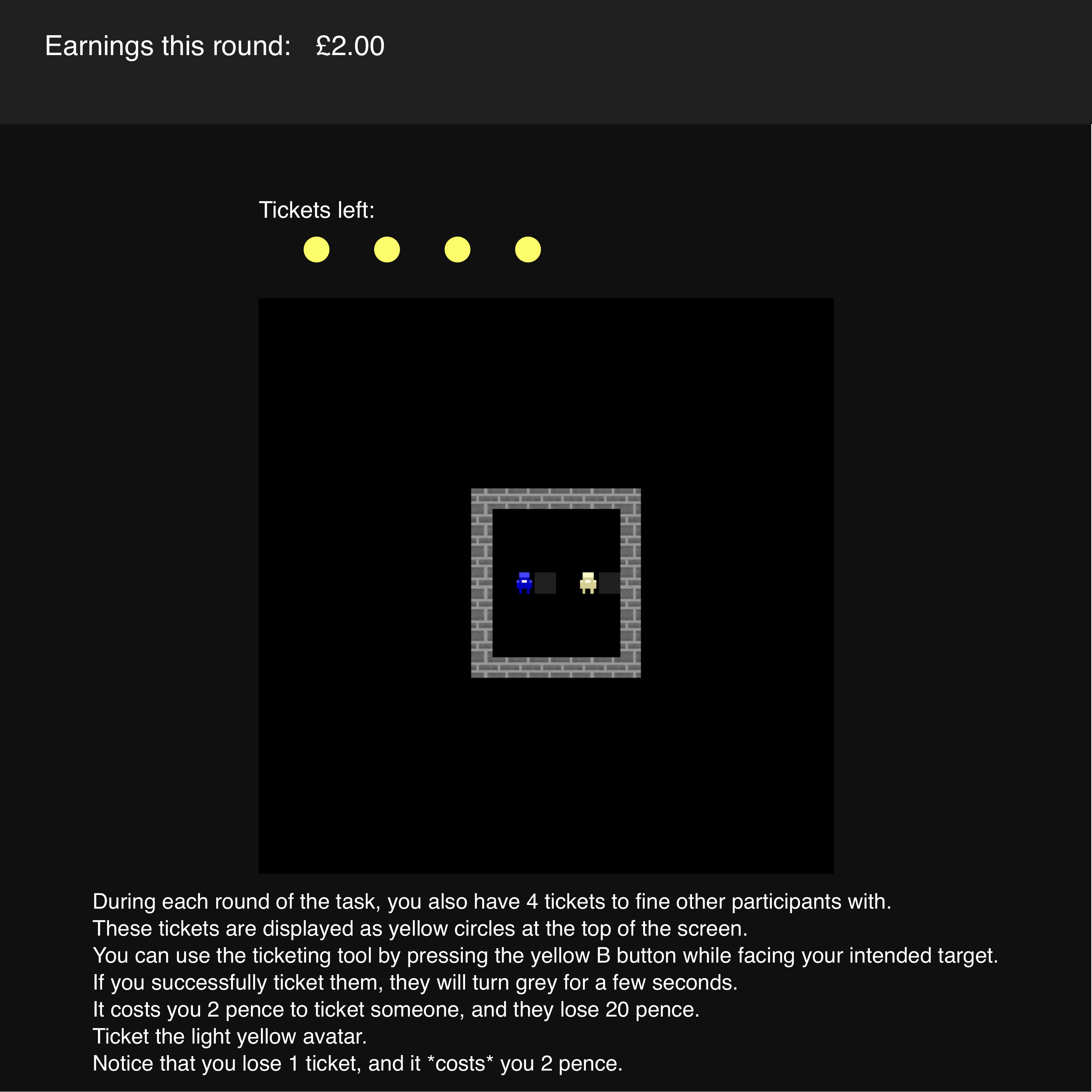}} \\
    \subfloat[(b)]{\includegraphics[width=10cm]{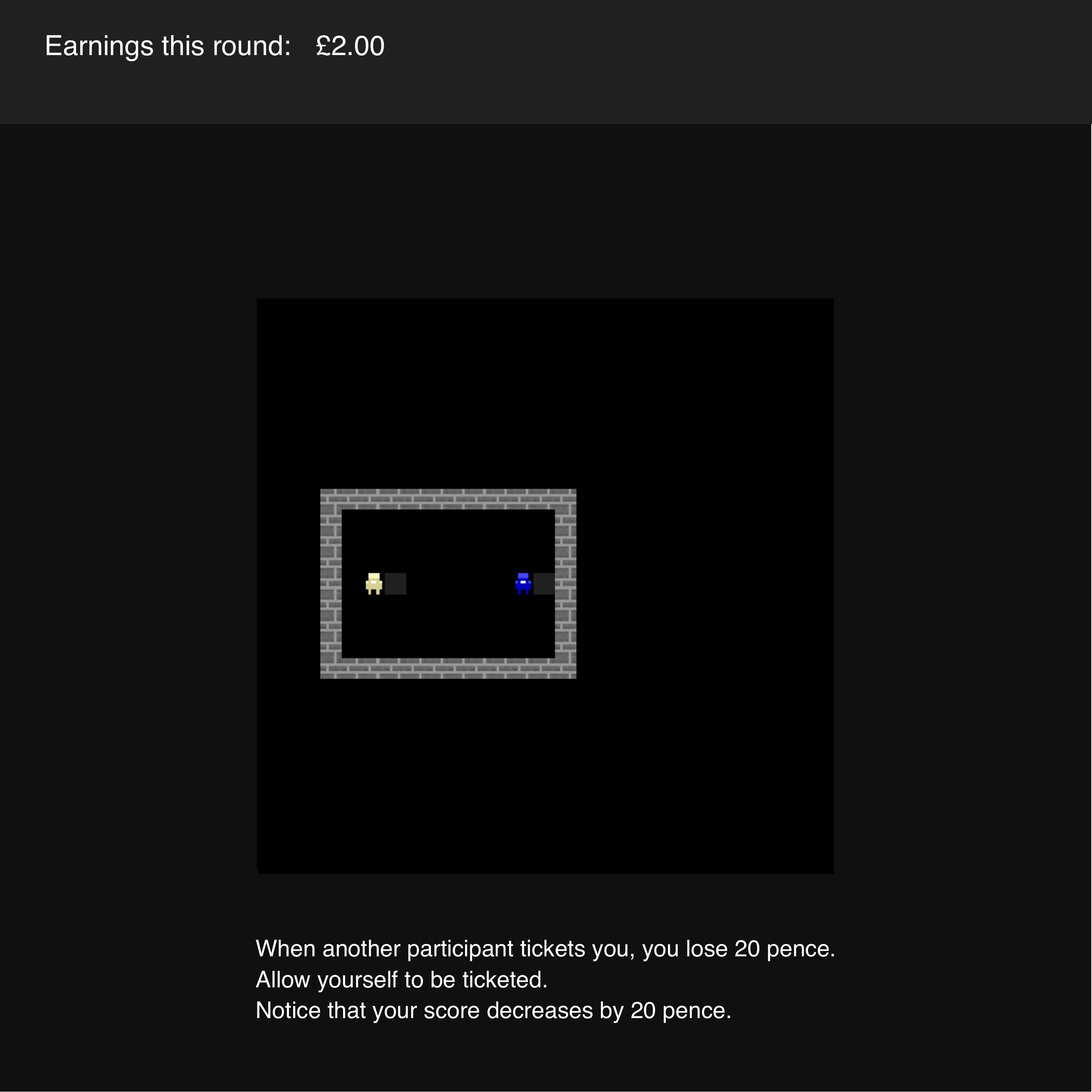}}
    \caption{Figure \ref*{fig:tutorials_e_f}a-b: Participants completed a number of tutorials to help them learn the controls for the task and the environmental dynamics of Clean Up.}
    \label{fig:tutorials_e_f}
\end{figure}

During the first stage of the experiment, participants receive a series of six tutorials on the action controls and the environmental dynamics for the Clean Up task (Figures \ref{fig:tutorials_a_b}-\ref{fig:tutorials_e_f}). The tutorials aimed to familiarize participants with (1) avatar movement, (2) apple collection, (3) pollution accumulation and the cleaning tool, (4) the effects of pollution on apple growth, (5) the ticketing tool and the cost of giving a ticket, and (6) the cost of receiving a ticket. Participants were subsequently instructed on the group nature of the task, including an explanation of the symmetry of information available about their own behavior to themselves and to their peers. Participants were also informed of the performance incentivization (i.e., the rules for receiving a bonus) at this stage. The tutorials described the river pollution as ``dirt'' to avoid explicitly priming participants with environmental concerns or pro-sustainability motives.

\begin{figure}[ht]
    \centering
    \subfloat[(a)]{\includegraphics[width=7.5cm]{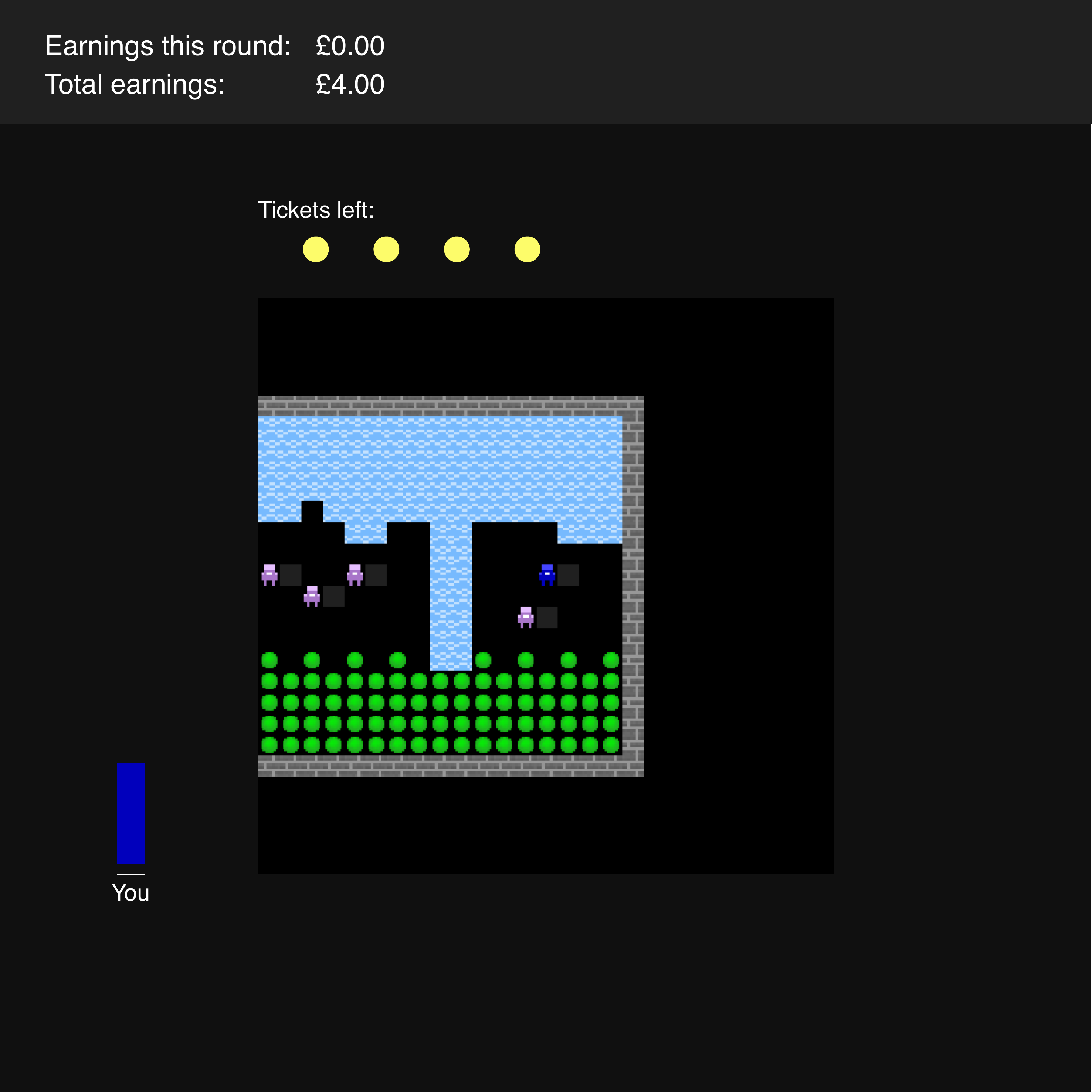}} \:
    \subfloat[(b)]{\includegraphics[width=7.5cm]{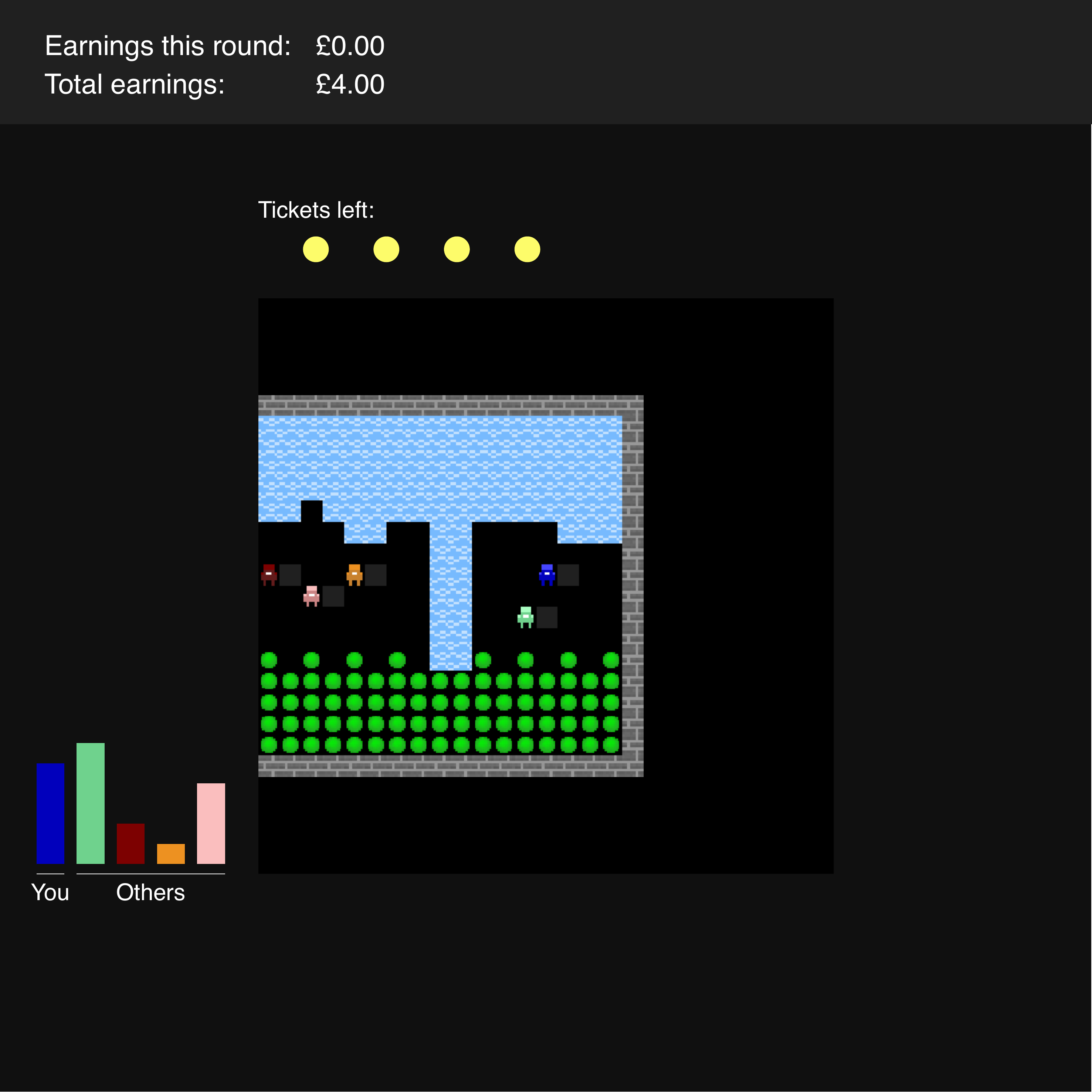}}
    \caption{Figure \ref*{fig:screenshot}: Participant view of the Clean Up task varied by condition. In both conditions, participants observed the environment through a 27 by 27 window, centered on their avatar (colored blue). Participants also observed their earnings for the current episode, their cumulative earnings through the current episode, and the number of tickets they had available. (a) In the anonymous condition, other participants were represented by lavender avatars. Each participant observed their own contribution level (the blue bar), but did not receive the contribution levels of others in their group. (b) In the identifiable condition, other participants were represented by uniquely colored avatars. Each participant observed their own contribution level, as well as the contribution levels of the others in their group (other colored bars).}
    \label{fig:screenshot}
\end{figure}

During episodes, participants observed the environment through a 27 by 27 window, centered around their avatar (Figure \ref{fig:screenshot}). The size of this observation window allows for a participant to view the entire map by standing in the middle of the map. However, practically speaking, participants spent the majority of the time playing with imperfect observability. In 94.3\% of steps, participants were positioned such that part of the map was obscured from their observation.

Episodes ran for $T=2000$ steps (approximately 2 minutes). Participants were not told the exact length of the episodes. At the end of every episode, each participant’s score for that episode was added to his or her cumulative score for the entire experiment. After the experiment, participants were paid a base payment of £15 and a bonus for their cumulative score at the rate of 0.5 pence per point, up to a maximum bonus of £30.

The University College London Research Ethics Committee conducted ethical review for the project and approved the study protocol (CPB/2013/015). All participants provided informed consent for the study.

The experiment was completed by 120 participants (age: mean $m = 21.5$, standard deviation $sd = 2.3$; gender: 50 male, 70 female), drawn from the University College London psychology department participant pool.\footnote{During one experimental session (i.e., ten participants in the lab for the experiment), one participant’s computer malfunctioned for several episodes. Participants in the dropped session were paid and debriefed as normal. To maintain a balanced design, the entire session was dropped from data analysis and an additional session was scheduled to fulfill the original design. Counting this dropped session, 130 participants were recruited across 13 sessions.} Participants earned an average of £30.85 ($sd =$ £6.44) during the experiment.

\section{Social Dilemma Analysis} \label{sec:social_dilemma_analysis}

Social dilemmas are situations in which there is a tradeoff between short-term individual incentives and long-term collective interest \cite{hughes2018inequity, kollock1998social, rapoport1974prisoner}. In this paper, we study public goods dilemmas, a particular subset of social dilemmas.

In the canonical public goods task \cite{fehr2000cooperation}, $n$ participants receive an initial endowment of $e$ tokens and choose a contribution level $c \in \{0, \dots, e\}$. Subsequently, contributions are pooled and increased by multiplication factor $M$, forming the public good, $G$:
\begin{equation}
   G = M \sum_{j = 1}^{n}{c_j} \label{public_goods_formula} \, .
\end{equation}
The public good is then distributed evenly across all $n$ participants. The collective group payoff (i.e., environmental reward) is consequently described by:
\begin{equation}
   U_{\textrm{total}} = n \cdot e + (M - 1) \sum_{j = 1}^{n}{c_j} \, , \label{group_payoff_formula}
\end{equation}
where the payoff to participant $k$ is described by:
\begin{equation}
   u_k = e - c_k + \frac{M}{n} \sum_{j = 1}^{n}{c_j} \, . \label{individual_payoff_formula}
\end{equation}

In the canonical setting, there is a deterministic relationship between the group's contributions, the size of the public good, and the group payoff. Similarly, the equal distribution of the public good deterministically defines the payoffs for individual participants. As a result, in the traditional public goods task, we can purposefully instantiate a social dilemma---a situation in which there is a tradeoff between individual incentives and the collective interest---by requiring $1 < M < n$.

\begin{figure}[ht]
    \centering
    \subfloat[(a)]{\includegraphics[width=6cm]{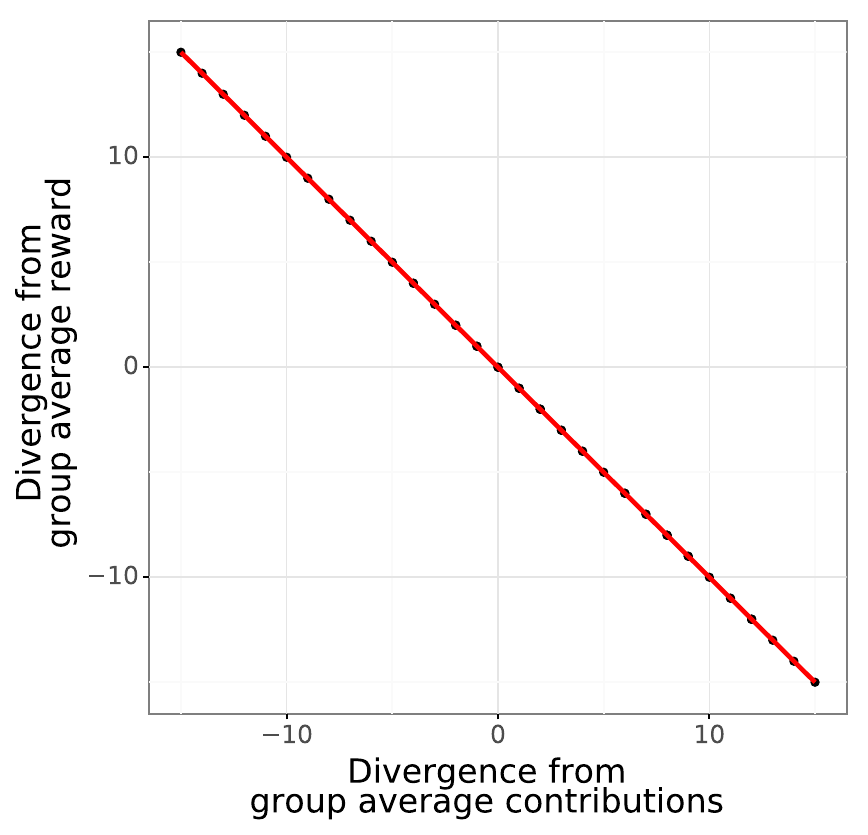}}
    \subfloat[(b)]{\includegraphics[width=6cm]{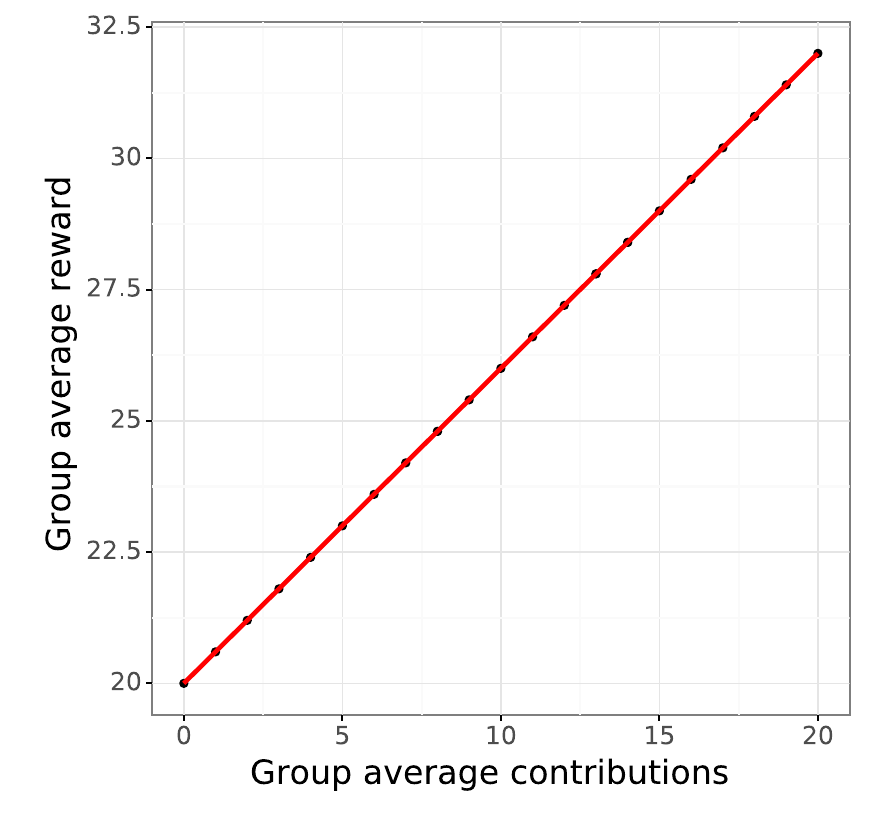}}
    \caption{Figure \ref*{fig:canonical_dilemma}: The incentive structure of the canonical public goods task \cite{fischbacher2001people} exhibits a distinctive sign reversal between the individual and group level. Contribution level is negatively correlated with reward within groups, but positively correlated between groups. (a) Participants who contribute less relative to their group receive higher reward. (b) However, groups that contribute more earn more reward.}
    \label{fig:canonical_dilemma}
\end{figure}

In the traditional parameterization of the public goods task, $n = 4$, $e = 20$, and $M = 1.6$ \cite{fehr2000cooperation}. With these parameters, the payoff functions \ref{group_payoff_formula} and \ref{individual_payoff_formula} respectively reflect a negative relationship between own contributions and own payoff, \textit{ceteris paribus}, and a positive relationship between group contributions and group payoff (Figure \ref{fig:canonical_dilemma}). This sign reversal is the defining characteristic of individual- and group-level incentive structures for social dilemmas.\footnote{The sign reversal between the individual and group level is conceptually related to Simpson's paradox, a well-studied statistical phenomenon whereby the direction of an association at the population-level reverses within the subgroups comprising that population (as described in \cite{simpson1951interpretation}; see also \cite{kievit2013simpson, pearson1899vi}).}

In stochastic, temporally and spatially extended tasks (cf., \cite{littman1994markov}), it is not straightforward to instantiate social-dilemma incentive structures with environmental rules \cite{leibo2017multi}. In our work, given the variation in environmental parameters between the human behavioral experiment and the model, it was especially important to confirm that both environment versions constituted a social dilemma.

Using linear regression, we characterize the extent to which empirical data match the sign-reversal pattern observed in the canonical case---what we might call a \textit{linear social dilemma} structure (cf., \cite{zelmer2003linear}). We conduct this analysis with two separate regressions. The first examines the individual-level incentive structure experienced by group members. The second examines the effects of collective behavior on group welfare. To match a linear social dilemma structure, we expect to see a negative relationship between cooperation and payoff at the individual level, but a positive relationship between cooperation and welfare at the group level.

\begin{gather}
   Y_j = \beta_0 + \beta_1 \cdot (c_j - \bar{c}) + \epsilon \, \\
   Y_g = \beta_0 + \beta_1 \cdot \bar{c} + \epsilon \, .
\end{gather}

In addition to this linear-dilemma analysis, we follow recent explorations in the multi-agent field by producing Schelling diagrams \cite{perolat2017multi, hughes2018inequity}. Schelling diagrams are an empirical approach to Markov games which characterize payoffs based on group policy composition. Visualizing payoff structures in this way has various benefits, including the opportunity to inspect whether the task reflects a social dilemma and the ability to identify game-theoretic concepts such as Nash equilibria and Pareto-optimal outcomes \cite{schelling1973hockey}. Schelling diagrams rely upon the ability to dichotomize the policy space (e.g., into cooperation and defection policies), as well as the ability to categorize individuals based on their observed behavioral trajectories.

Given the well-established use of punishment to decrease the payoffs of free riders \cite{henrich2006costly, gachter2008long}, for these analyses we use apple consumption to model payoff and welfare.

\subsection{Human Behavioral Experiment}

\begin{figure}[ht]
    \centering
    \subfloat[(a)]{\includegraphics[width=6cm]{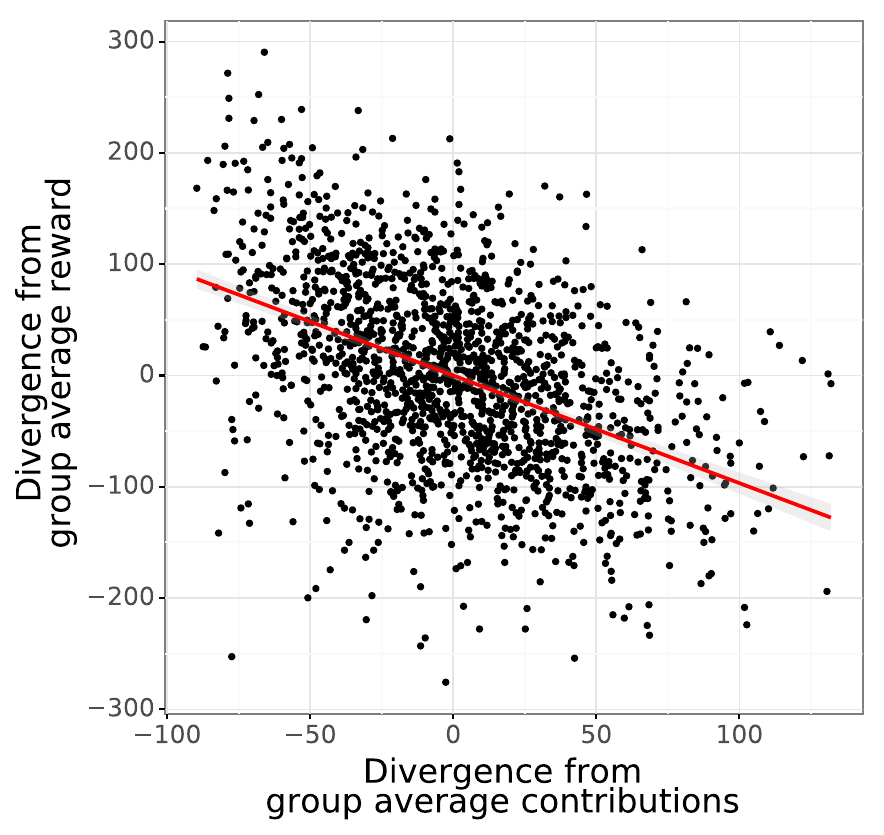} \label{fig:empirical_dilemma/a}}
    \subfloat[(b)]{\includegraphics[width=6cm]{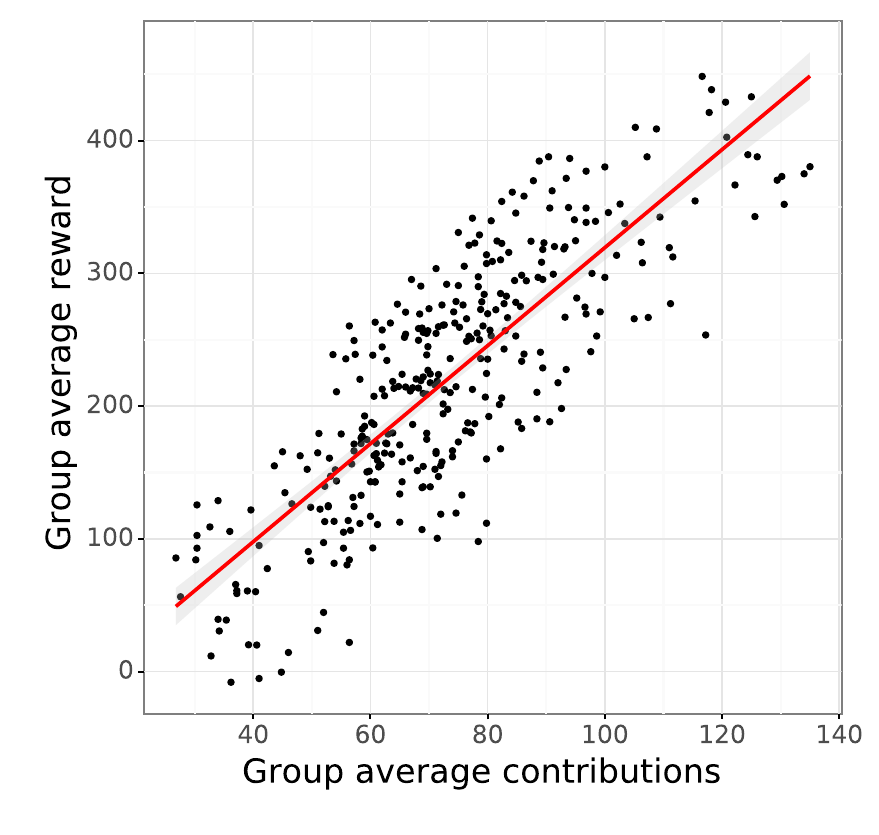} \label{fig:empirical_dilemma/b}}
    \caption{Figure \ref*{fig:empirical_dilemma}: Empirical inspection reveals a linear social dilemma structure for humans completing the Clean Up task. Contribution level is negatively correlated with reward within groups, but positively correlated between groups.  Errors bands reflect 95\% confidence intervals. (a) Participants who contribute less relative to their group receive higher reward. (b) However, groups that contribute more generate higher welfare.}
    \label{fig:empirical_dilemma}
\end{figure}

At the individual level, the amount that a participant's contributions exceeded or fell below their group's average contribution level had a significant negative relationship with the amount their payoff exceeded or fell below the group average payoff, $\beta = -0.97$, 95\% CI $[-1.06, -0.88]$, $p < 0.0001$ (Figure \ref{fig:empirical_dilemma/a}). Participants who contributed more relative to their peers collected fewer apples than their peers.

At the group level, average contribution level was significantly and positively associated with average apple consumption, $\beta = 3.69$, 95\% CI $[3.42, 3.97]$, $p < 0.0001$ (Figure \ref{fig:empirical_dilemma/b}). In contrast with the negative individual-level association, groups that collectively contribute more also collectively consume more apples.

Taken together, these two effects support the existence of a linear social dilemma incentive structure within Clean Up.

\begin{figure}[ht]
    \centering
    \includegraphics[width=8.75cm]{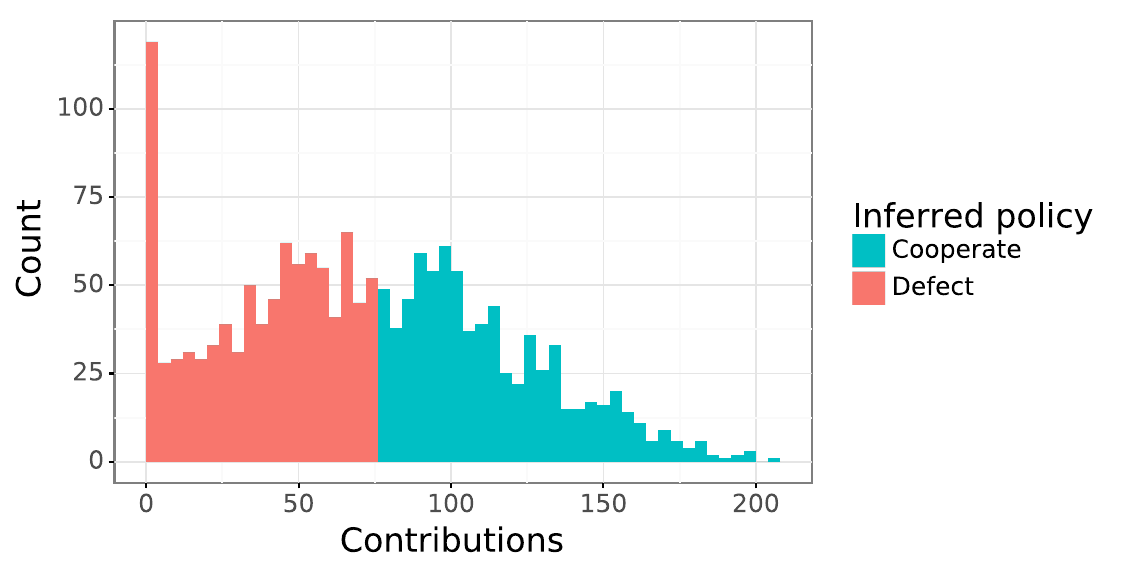}
    \caption{Figure \ref*{fig:jenks_contributions}: Individual contribution levels were variable across participants and episodes. We use the Jenks natural breaks method to dichotomize this distribution. This results in the categorization of contribution levels below 76 as ``defect'' policies and contribution levels at or above 76 as ``cooperate'' policies.}
    \label{fig:jenks_contributions}
\end{figure}

We sought to verify this finding by generating an empirical Schelling diagram for the human behavioral experiment. In effect, we map empirical data onto a policy space by categorizing observed behavioral trajectories into ``cooperate'' trajectories and ``defect'' trajectories. We chose to dichotomize participant contribution levels using the Jenks optimization method. This method identifies cutoff points which minimize within-category variance and maximize between-category variance, given a number of categories to establish \cite{jenks1967data}. The Jenks method identifies 76 contribution steps as the natural breakpoint dichotomizing the distribution of participant contribution levels (Figure \ref{fig:jenks_contributions}).

\begin{figure}[ht]
    \centering
    \includegraphics[width=8.75cm]{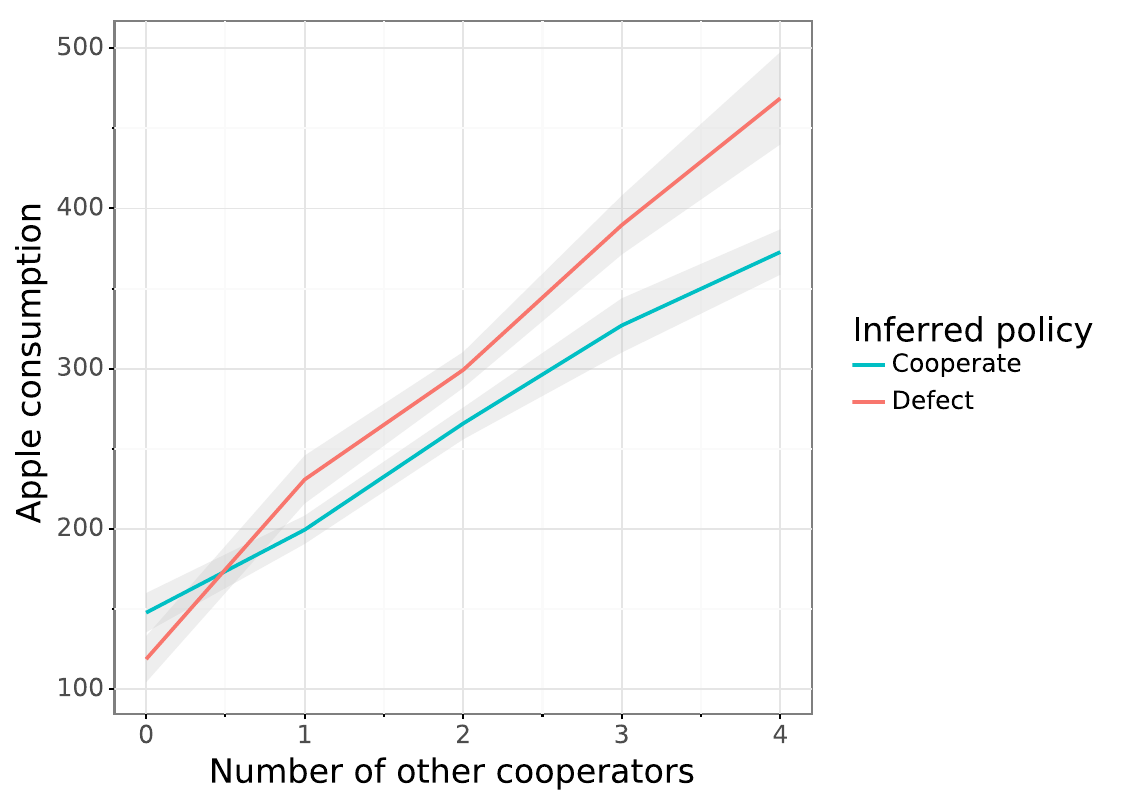}
    \caption{Figure \ref*{fig:human_schelling}: An inspection of the empirical Schelling diagram indicates that the Clean Up tasks meets Hughes and colleagues' \cite{hughes2018inequity} definition of a social dilemma.  Errors bands reflect 95\% confidence intervals.}
    \label{fig:human_schelling}
\end{figure}

We classified participants as cooperating in a given episode if they contributed more than this threshold and as defecting if they contributed less. Subsequently, we tabulated the number of cooperators and defectors in each episode for every group. As before, we examine apple consumption as the relevant payoff for participants. Average apple consumption is calculated separately among cooperators and defectors for each episode. To construct the Schelling diagram, average apple consumption for cooperators and defectors is plotted against the count of cooperators in each episode (Figure \ref{fig:human_schelling}).

Hughes and colleagues \cite{hughes2018inequity} delineate the following conditions to define a binary-choice social dilemma:

\begin{enumerate}
    \item Mutual cooperation is preferred over mutual defection: $R_c(N) > R_d(0)$.
    \item Mutual cooperation is preferred to being exploited by defectors: $R_c(N) > R_c(0)$.
    \item Either...
    \begin{enumerate}
        \item Mutual defection is preferred to being exploited (fear): $R_d(i) > R_c(i)$ for sufficiently small $i$,
        \item Or exploiting a cooperator is preferred to mutual cooperation (greed). $R_d(i) > R_c(i)$ for sufficiently large $i$.
    \end{enumerate}
\end{enumerate}

This definition can be translated to a frequentist framework by instantiating each of these conditions with a one-sided independent \textit{t}-test. These \textit{t}-tests produce the following results with the experimental Clean Up data:

\begin{itemize}
    \item Condition 1 is met. The payoff to cooperators under mutual cooperation ($m = 372.2$, $sd = 35.23$) is significantly higher than the payoff to defectors under mutual defection ($m = 109.6$, $sd = 28.6$), $t(36.4) = 25.9$, $p_1 < 0.0001$.
    \item Condition 2 is met. The payoff to cooperators under mutual cooperation ($m = 372.2$, $sd = 35.23$) is significantly higher than the payoff to cooperators when all other group members defect ($m = 150.0$, $sd = 56.7$), $t(63.4) = 22.8$, $p_2 < 0.0001$.
    \item Condition 3a is not met. The payoff to defectors under mutual defection ($m = 109.6$, $sd = 28.6$) is not significantly higher than the payoff to cooperators when all other group members defect ($m = 150.0$, $sd = 56.7$), $t(44.4) = -4.2$, $p_{3a} = 1.00$.
    \item Condition 3b is met. The payoff to defectors when all other group members cooperate ($m = 465.5$, $sd = 84.8$) is higher than the payoff to cooperators under mutual cooperation ($m = 372.2$, $sd = 35.23$), $t(43.8) = 5.6$, $p_{3b} < 0.0001$.
\end{itemize}

We synthesize the results of these \textit{t}-tests ($p_1$, $p_2$, $p_{3a}$, $p_{3b}$, respectively) into a single statistical test through two steps. First, we use Fisher's method \cite{fisher1925statistical} to consolidate $p_{3a}$ and $p_{3b}$ while controlling for multiple comparisons. This results in the joint \textit{p}-value $p_3$. Second, we use a maximum $p$-value approach to combine $p_1$, $p_2$, and $p_3$: $p_{\textrm{overall}} = \max \left(p_1, p_2, p_3\right)$.

In the first step, Fisher's method indicates a significant overall result for condition 3, $\chi^2(4) = 30.3$, $p_3 < 0.0001$. In the second step, combining conditions 1, 2, and 3 results in $p_{\textrm{overall}} < 0.0001$. According to this combined significance test, the Clean Up task meets Hughes and colleagues' definition of a social dilemma \cite{hughes2018inequity}.

\subsection{Computational Model}

\begin{figure}[ht]
    \centering
    \includegraphics[width=8.75cm]{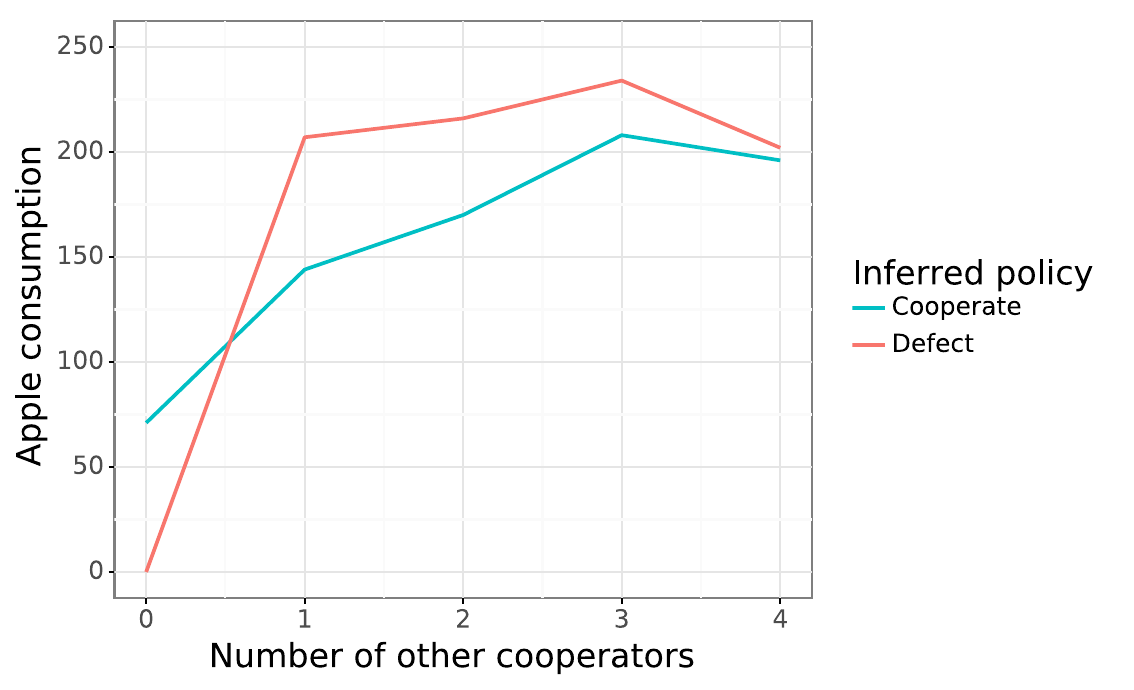}
    \caption{Figure \ref*{fig:rl_schelling}: Previous work with reinforcement learning agents \cite{hughes2018inequity} demonstrated that the Clean Up task has a social dilemma incentive structure. Reproduced with permission from \cite{hughes2018inequity}.}
    \label{fig:rl_schelling}
\end{figure}

Previous work by Hughes and colleagues \cite{hughes2018inequity} verified that the incentive structure of Clean Up produces social dilemma pressures for reinforcement learning agents (Figure \ref{fig:rl_schelling}). In that work, agents were trained with a specialized protocol. During training, the ability to contribute to the public good was withheld from some agents; a small group reward signal was added to the remaining agents. The former and latter types of agent were classified as defectors and cooperators, respectively. As a result, groups varied in their composition of cooperating and defecting policies. Echoing our findings with the human behavioral analysis, Hughes and colleagues found that the resulting empirical Schelling diagram matched condition 1 ($R_c[4] > R_d[0]$), condition 2 ($R_c[4] > R_c[0]$) and condition 3b ($R_d[4] > R_c[4]$).

Presaging our findings for the human participants, Hughes and colleagues concluded from this pattern that the Clean Up task instantiates a social dilemma for reinforcement learning agents.

\section{Comprehension Check}
\subsection{Human Behavioral Experiment}

Two open-ended questions at the end of the post-experiment questionnaire asked participants whether they attended to the bar display during the episodes when there was one bar and during the episodes when there were five bars. Responses were coded dichotomously as \textit{no, did not attend} or \textit{yes, attended} (Table \ref{tab:mcnemar_manip_check}).

\begin{table}[ht]
    \centering
    \begin{tabular}{c c|c c}
     & \multicolumn{1}{c}{} & \multicolumn{2}{c}{Attended to the} \\
     & \multicolumn{1}{c}{} & \multicolumn{2}{c}{1-bar display?} \\
     & & \multicolumn{1}{c}{Yes} & \multicolumn{1}{c}{No} \\ 
    \cline{2-4}
    Attended to the & \multicolumn{1}{r|}{Yes} & 39 & 74 \\  
    5-bar display? & \multicolumn{1}{r|}{No} & 0 & 7    
    \end{tabular}
    \caption{Table \ref*{tab:mcnemar_manip_check}: Frequency of participant \textit{Yes} and \textit{No} responses to the comprehension check question in the post-experiment questionnaire.}
    \label{tab:mcnemar_manip_check}
\end{table}

An exact McNemar’s test indicates a statistically significant difference in the proportion of participants who reported attending to the 1-bar display and the proportion who reported attending to the 5-bar display, $\chi^2 = 74$, $p < 0.0001$. A significantly greater proportion of participants reported attending to the 5-bar display (94.2\%) than reported attending to the 1-bar display (32.5\%).

\begin{figure}[ht]
    \centering
    \subfloat[(a)]{\includegraphics[width=8cm]{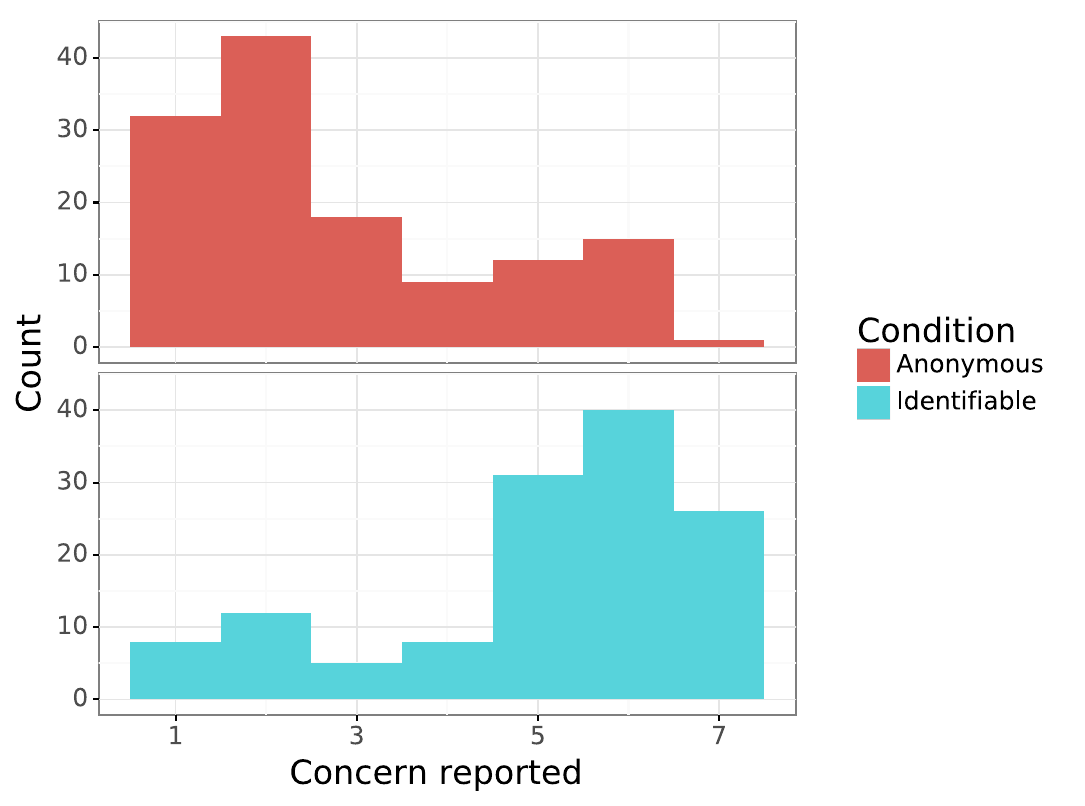}}
    \subfloat[(b)]{\includegraphics[width=6cm]{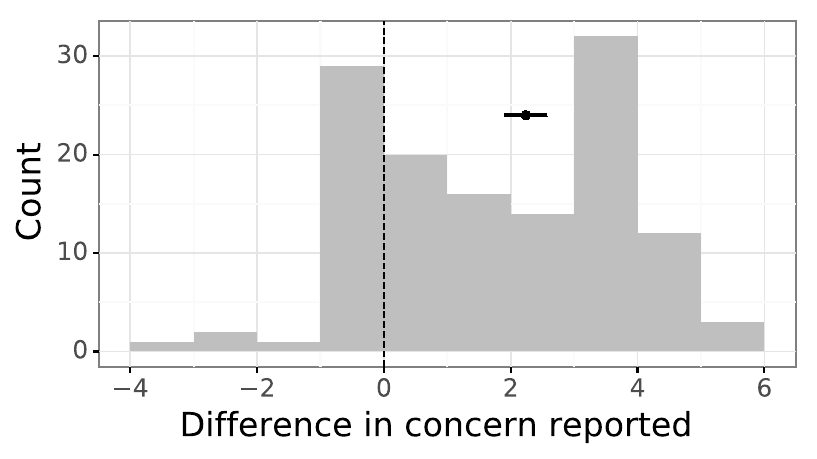}}
    \caption{Figure \ref*{fig:concern_manip_check}: (a) Distributions of participant agreement with the statement ``I was concerned with what other participants thought about how much dirt I was cleaning up'' varied by condition. (b) Participants were significantly more likely to report such concern in the identifiable condition than in the anonymous condition. The plotted point reflects the sample mean difference ($m_{\textrm{diff}} = 2.21$), with error bars reflecting the 95\% confidence interval.}
    \label{fig:concern_manip_check}
\end{figure}

Using two 7-point scales, the questionnaire additionally assessed the degree to which participants were concerned about others tracking their behavior during each condition. Participant concern scores were centered on $m = 4.03$ ($sd = 1.81$) in the identifiable condition and on $m = 1.82$ ($sd = 1.71$) in the anonymous condition (Figure \ref{fig:concern_manip_check}). Participants reported being significantly more concerned about others tracking their contribution behavior in the identifiable condition than in the anonymous condition, $m_{\textrm{diff}} = 2.21$ ($95\%$ CI $= [1.85, 2.57]$), $t(119) = 12.25$, $p < 0.0001$ (Figure \ref{fig:concern_manip_check}).

\section{Main Group Effects} \label{sec:main_effects}
\subsection{Computational Model}

The reinforcement learning agents in the computational model did not update their policy during the evaluation stage of the experiment. As a result, the order in which groups experienced the conditions did not influence their behavior. We consequently evaluate the effect of condition on each social outcome metric $Y_g$ with a one-way, repeated-measure analysis of variance (ANOVA):

\begin{equation} \label{eqn:agent_anova}
   Y_g = \beta_0 + \beta_1 \cdot \textrm{Condition} + \mu_g + \epsilon \, .
\end{equation}

We first conduct a one-way ANOVA for collective return. There was a significant effect of condition on collective return, $F(1,311) = 2030.3$, $p < 0.0001$. In the model, groups earned significantly more in the identifiable condition (604 points on average) than in the anonymous condition (339 points on average).

We next conduct a one-way ANOVA for group contribution level. There was a significant effect of condition on group contribution level, $F(1,311) = 1090.7$, $p < 0.0001$. In the model, groups cleaned significantly more in the identifiable condition (for 422 steps on average) in the identifiable condition than in the anonymous condition (299 steps on average).

\subsection{Human Behavioral Experiment}

The human behavioral experiment took a counterbalanced, within-participant design: each participant was exposed to all experimental conditions, and the ordering of conditions was balanced across participants. Half of the participant groups completed the identifiable condition first and the anonymous condition second, and the other half completed the anonymous condition first and the identifiable condition second.

We would like to evaluate the effect of condition (identifiability versus anonymity) on several outcome metrics. In the human behavioral experiment, the counterbalanced design indicates the use of two-way ANOVA. To facilitate comparison with the model results, in the main text we report the main effects from the two-way ANOVA. Here we present the full models, including the main effect of task and the interaction effect. For this experiment, a two-way, repeated-measures ANOVA was used to assess the effect of task condition (identifiability or anonymity) and task number (first task or second task) on each social outcome metric $Y_g$:

\begin{equation}
   Y_g = \beta_0 + \beta_1 \cdot \textrm{Condition} + \beta_2 \cdot \textrm{Task} + \beta_3 \cdot \textrm{Condition} \times \textrm{Task} + \mu_g + \epsilon \, .
\end{equation}

\begin{figure}[ht]
    \centering
    \includegraphics[width=8cm]{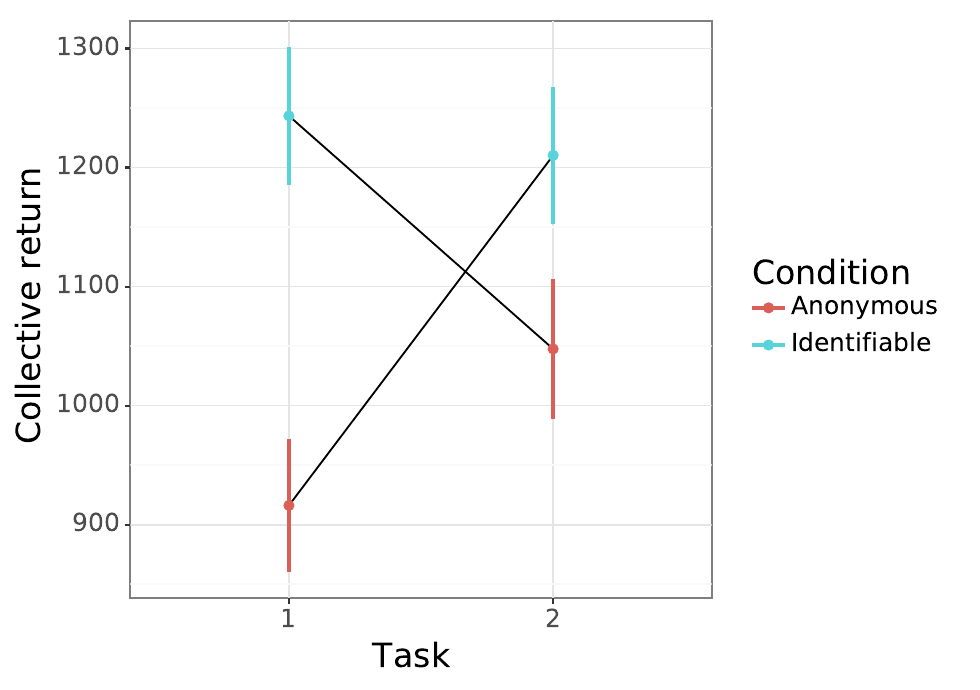}
    \caption{Figure \ref*{fig:anova_collective_return}: We use a two-way, repeated-measures ANOVA to evaluate the effects of condition and task number on collective return. As expected, there was a significant main effect of condition on collective return. The main effect of task number and the interaction effect between condition and task number were not significant. Error bars reflect 95\% confidence intervals.}
    \label{fig:anova_collective_return}
\end{figure}

We first conduct a two-way ANOVA for collective return (Figure \ref{fig:anova_collective_return}). There was a significant main effect of condition on collective return, $F(1,310) = 89.4$, $p < 0.0001$. The main effect of task number was non-significant, $F(1,310) = 3.6$, $p = 0.059$. The interaction effect was also non-significant, $F(1,22) = 0.2, p = 0.63$. Groups earned significantly more in the identifiable condition (1227 points on average) than in the anonymous condition (982 points on average).

\begin{figure}[ht]
    \centering
    \includegraphics[width=8cm]{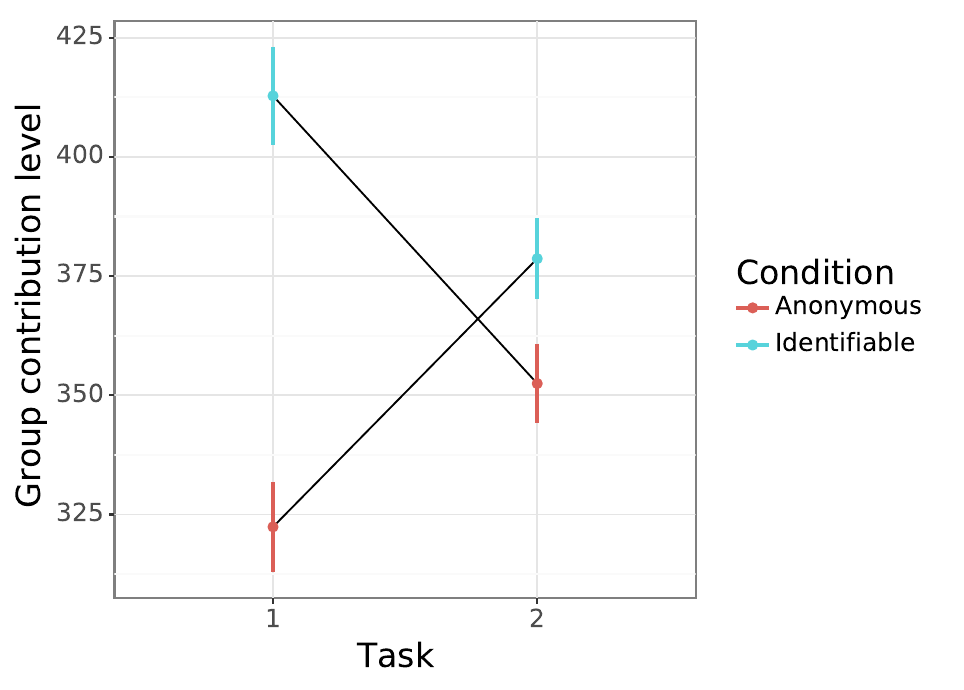}
    \caption{Figure \ref*{fig:anova_contributions}: We use a two-way, repeated-measures ANOVA to evaluate the effects of condition and task number on group contribution level. As expected, there was a significant main effect of condition on group contribution level. The main effect of task number and the interaction effect between condition and task number were not significant. Error bars reflect 95\% confidence intervals.}
    \label{fig:anova_contributions}
\end{figure}

We next conduct a two-way ANOVA for total contribution level (Figure \ref{fig:anova_contributions}). There was a significant main effect of condition on total contribution level, $F(1,310) = 199.4$, $p < 0.0001$. The main effect of task number was non-significant, $F(1,310) = 0.2$, $p = 0.62$. The interaction effect was also non-significant, $F(1,22) = 0.67$, $p = 0.42$. Groups cleaned significantly more in the identifiable condition (for 396 steps on average) in the identifiable condition than in the anonymous condition (337 steps on average).

\section{Spatial Coordination Analysis}

To explore the use of spatial coordination strategies in Clean Up, we estimate the extent of territoriality groups used to coordinate their investments in the public good. Our measure of territoriality relies on beta diversity, a measure of compositional heterogeneity across locations originally developed by ecologists \cite{koleff2003measuring, whittaker1960vegetation}. Conceptually, we calculate a metric that communicates the degree to which group members' ``territories'' overlapped with each other. We start by converting group member trajectories to presence-absence data for each location in the map. Each location $l_k$ that is visited at least once is recorded in a vector $\mathbf{l} = \{l_1, ... l_{N_l}\}$, of length $N_l$, where $N_l$ denotes the number of distinct locations that were visited. Each group member $j$ that visits location $l_k$ is also recorded, resulting in a corresponding vector $\mathbf{j}_{l_k}$ of length $n_{l_k}$, where $n_{l_k}$ denotes the number of distinct group members who visited location $l_k$. We use the presence-absence data for group members' movements within the river region to calculate alpha, gamma, and beta diversities. Alpha diversity is defined as the number of group members who were present at the average location:
\begin{equation}
   \alpha_d = \frac{1}{N_l} \sum_{l_k \in \pmb{l}}{n_{l_k}} \, .
\end{equation}
Gamma diversity is defined as the number of unique group members who were present over all locations:
\begin{equation}
   \gamma_d = \left| \bigcup_{l_k \in \pmb{l}} \pmb{j}_{l_k} \right| \, .
\end{equation}
Beta diversity, the effective number of different group compositions, is defined as the ratio of gamma to alpha diversity:
\begin{equation}
   \beta_d = \frac{\gamma_d}{\alpha_d} \, .
\end{equation}

Beta diversity is lower bounded by $1$ (representing a single group composition for all visited locations) and upper bounded by whichever of $\gamma_d$ and $N_l$ is lower (representing completely non-overlapping locations visited by group members or completely different compositions per location, respectively). To account for the variability of the upper bound across episodes, we calculate a normalized beta diversity:
\begin{equation}
   {\beta_d}' = \frac{\beta_d}{\textrm{min}\left( \gamma_d, N_l \right)} \, .
\end{equation}

We use the normalized beta diversity as a measure of territoriality. A territoriality of 0 indicates that group members' territories were identical. A territoriality of 1, in contrast, indicates that group members' territories were entirely disjoint.

\subsection{Computational Model}

We conduct a one-way ANOVA to assess the effect of the intrinsic motivation for reputation on group territoriality. There was a significant effect of condition on territoriality, $F(1,311) = 432$, $p < 0.0001$. In the model, groups exhibited significantly less territoriality in the identifiable condition (with an average territoriality score of 0.42) than in the anonymous condition (with an average score of 0.58).

\subsection{Human Behavioral Experiment}

\begin{figure}[!t]
    \centering
    \includegraphics[width=8cm]{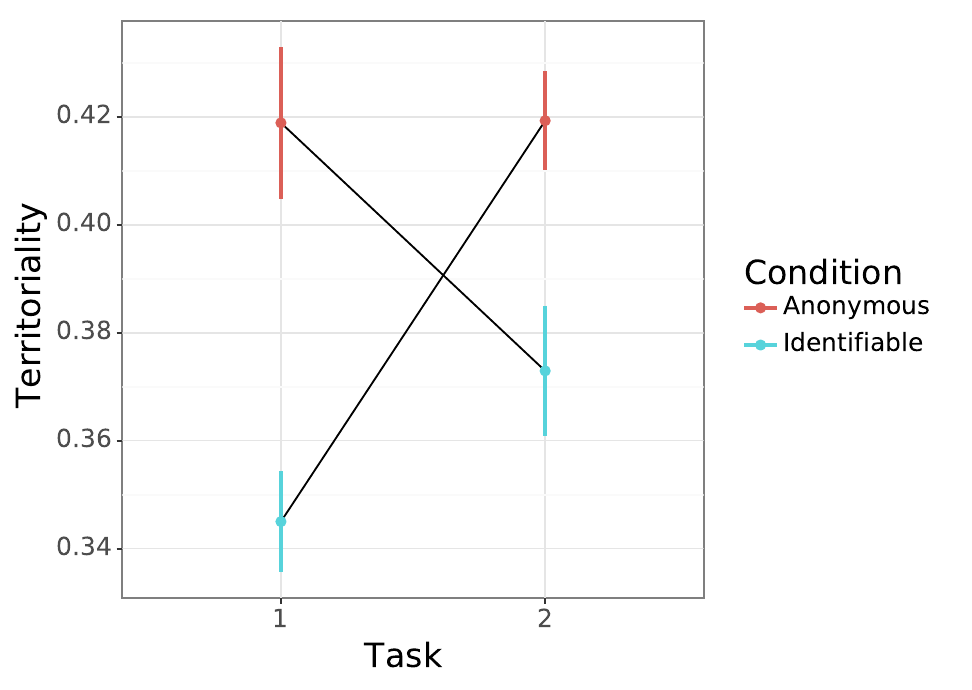}
    \caption{Figure \ref*{fig:anova_territoriality}: We use a two-way, repeated-measures ANOVA to evaluate the effects of condition and task number on group territoriality. There was a significant main effect of condition on group territoriality. The main effect of task number was also significant, while the interaction effect between condition and task number was not significant. Error bars reflect 95\% confidence intervals.}
    \label{fig:anova_territoriality}
\end{figure}

We conduct a two-way ANOVA to assess the effect of the intrinsic motivation for reputation on group territoriality (Figure \ref{fig:anova_territoriality}). As before, we highlight the main effect of condition in the main text to facilitate comparison with the model results, and here expand on the other terms of the two-way ANOVA. There was a significant main effect of condition on territoriality, $F(1,310) = 138.4$, $p < 0.0001$. The main effect of task number was also significant, $F(1,310) = 7.7$, $p = 0.0059$. The interaction effect was non-significant, $F(1,22) = 0.2$, $p = 0.67$. Groups were significantly less territorial in the identifiable condition (with an average territoriality score of 0.36) than in the anonymous condition (with an average score of 0.42).

\section{Temporal Coordination Analysis}

For this analysis, our aim is to understand whether groups organized their behavior using temporal coordination strategies. Toward this end, we develop {two measures of temporal coordination: a measure of group turn taking and a measure of temporal consistency for group contributions.}

\begin{table}[ht]
    \centering
    \begin{tabular}{c|c}
    Turns since group & Recency value\\
    member $j$'s last turn & for this turn \\
    \cline{1-2}
    $0$ & $1$ \\
    $1$ & $0.75$ \\
    $2$ & $0.50$ \\
    $3$ & $0.25$ \\
    $4+$ & $0$ \\ 
    \end{tabular}
    \caption{Table \ref*{tab:turn_recency}: Mapping between the number of turns that occurred since group member $j$'s last turn in the river and the assigned recency value.}
    \label{tab:turn_recency}
\end{table}

Turn taking is calculated based on the ordering of group member ``turns'' entering the river to clean pollution. For each episode, we record the sequence $\mathbf{S}$ of group members entering the river. For each of the identities in this sequence of turns, we calculated a recency value reflecting the number of turns that had occurred since the group member's last turn in the river (Table \ref{tab:turn_recency}). Turn-taking scores are generated by averaging the recency values for each turn in $\mathbf{S}$, taking the additive inverse of the average, and then adding a constant of 1 to the subsequent metric. A turn-taking score of 0 represents an episode where a single group member took all turns in the river. A turn-taking score of 1 reflects an episode where all group members rotated into the river, such that four turns pass between each of group member $j$'s turns in the river.

\begin{figure}[ht]
    \centering
    \includegraphics[width=8.75cm]{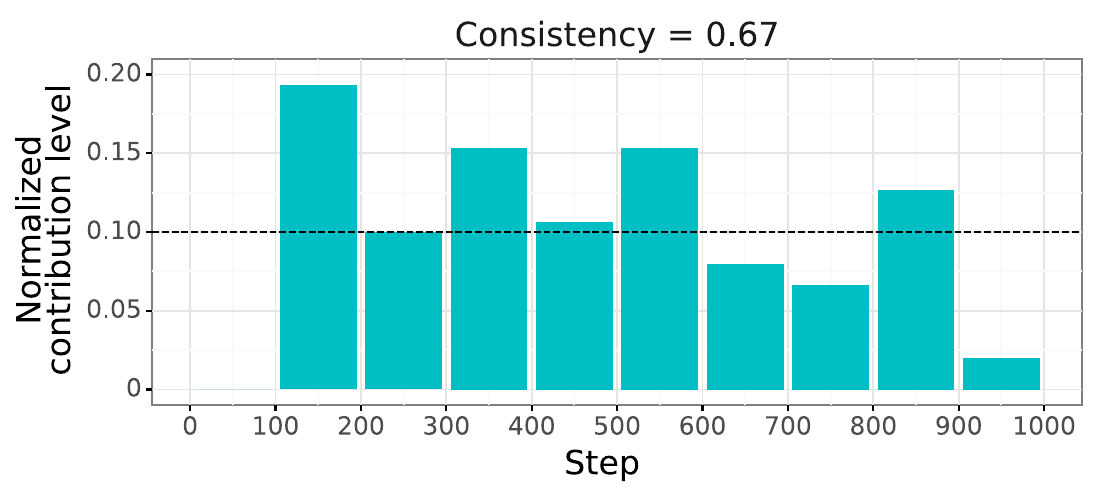} \\
    \includegraphics[width=8.75cm]{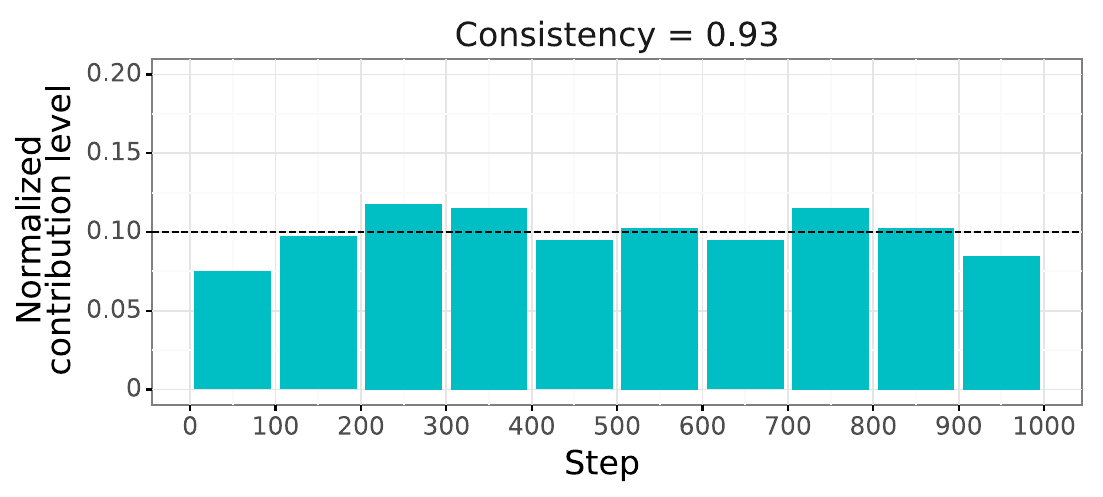}
    \caption{Figure \ref*{fig:temp_consis}: Example patterns of contribution density over time from the model, showing two groups from the computational model. In the first example, the group achieved low temporal consistency ($\textrm{consistency} = 0.67$). In the second, the group achieved high temporal consistency ($\textrm{consistency} = 0.93$).}
    \label{fig:temp_consis}
\end{figure}

Consistency is calculated by binning the full sequence of contributions from an episode into a number of granular periods. Here we estimate consistency using $t = 10$ periods for each episode (see Figure \ref{fig:temp_consis}).

Contributions within these periods are summed, forming a vector $\mathbf{c_T}$ of binned contributions. A consistency score is then calculated for this binned contribution vector, measuring equality over the temporal dimension:

\begin{equation}
   \textrm{Gini}(\mathbf{c_T}) = \frac{\displaystyle \sum_{j=1}^t \sum_{k=1}^t |c_j - c_k|}{\displaystyle 2 n^2 \bar{c}} \, ,
\end{equation}
\begin{equation}
   \label{eqn:consistency}
   \textrm{Consistency}(\mathbf{c_T}) = 1 - \textrm{Gini}(\mathbf{c_T}) \, .
\end{equation}

Groups that concentrate all of their contribution efforts in a short span of time exhibit low temporal consistency, whereas groups that evenly apportion their contribution efforts over time manifest high temporal consistency (Figure \ref{fig:temp_consis}). Maintaining a high level of contribution consistency requires a group to coordinate which members will clean the river at any given time.

\subsection{Computational Model}

We conduct a one-way ANOVA to assess the effect of the intrinsic motivation for reputation on group turn taking. There was a significant effect of condition on turn taking, $F(1,311) = 758.3$, $p < 0.0001$. In the model, groups exhibited significantly more turn taking in the identifiable condition (with an average turn-taking score of 0.75) than in the anonymous condition (with an average score of 0.56).

{In the main text, we evaluate the relationship between turn taking and group performance with a linear regression, averaging observations by group:}

\begin{equation}
   \textrm{Collective Return} = \beta_0 + \beta_1 \cdot \textrm{Turn-Taking Score} + \epsilon \, .
\end{equation}

{With this regression, the model predicted a significant relationship between turn taking and collective return, $\beta = 1030.3$, 95\% CI $[908.1, 1152.5]$, $p < 0.0001$. Turn taking was positively associated with collective return, such that groups that relied more heavily on a rotation scheme tended to achieve higher scores.}

\begin{figure}[ht]
    \centering
    \includegraphics[width=11cm]{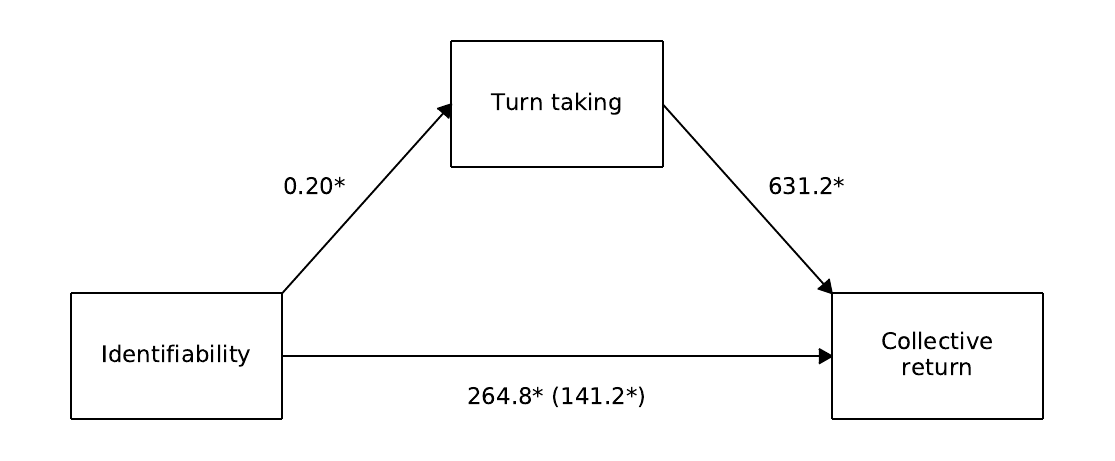}
    \caption{Figure \ref*{fig:agent_turn_taking}: Mediation analysis revealed a significant indirect effect of identifiability on collective return, mediated by group turn taking. * $p < 0.05$.}
    \label{fig:agent_turn_taking}
\end{figure}

{We conduct a mediation analysis \cite{rucker2011mediation} to estimate the indirect effect of identifiability on collective return through group turn-taking (Figure \ref{fig:agent_turn_taking}). The analysis reveals a significant and positive indirect effect of identifiability on collective return through group turn-taking, $AB = 123.6$, 95\% CI $[110.7, 138.7]$, $p < 0.0001$. Furthermore, the positive association between identifiability and collective return ($C = 264.8$, 95\% CI $[248.4, 279.4]$, $p < 0.0001$) is reduced after accounting for turn taking ($C' = 141.2$, 95\% CI $[110.7, 155.4]$, $p < 0.0001$).}

\begin{figure}[t]
    \centering
    \includegraphics[height=6cm]{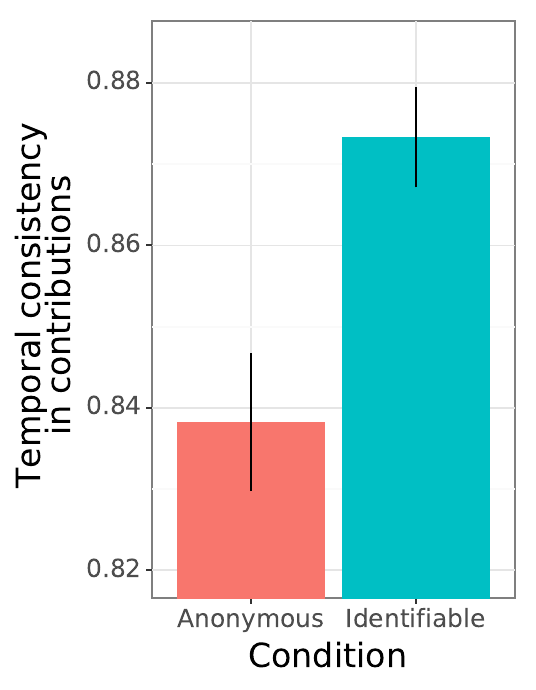}
    \caption{Figure \ref*{fig:agent_anova_consistency}: We use a repeated-measures ANOVA to evaluate the effect of condition on the temporal consistency of group contributions. Condition exerted a significant effect on contribution consistency. Error bars reflect 95\% confidence intervals.}
    \label{fig:agent_anova_consistency}
\end{figure}

{To further test our findings, we replicate these analyses with the temporal consistency measure. We conduct a repeated-measures ANOVA to assess the effect of the intrinsic motivation for reputation on group contribution consistency over time. Condition exerted a significant effect on temporal consistency, $F(1,311) = 81.1$, $p < 0.0001$ (Figure \ref{fig:agent_anova_consistency}). In the model, groups acted with greater consistency in the identifiable condition (with an average consistency score of 0.87) than in the anonymous condition (with an average score of 0.84).}

{We next evaluate the relationship between temporal consistency and group performance with a linear regression, averaging observations by group:}

\begin{equation}
   \textrm{Collective Return} = \beta_0 + \beta_1 \cdot \textrm{Contribution Consistency} + \epsilon \, .
\end{equation}

\begin{figure}[ht]
    \centering
    \includegraphics[width=6cm]{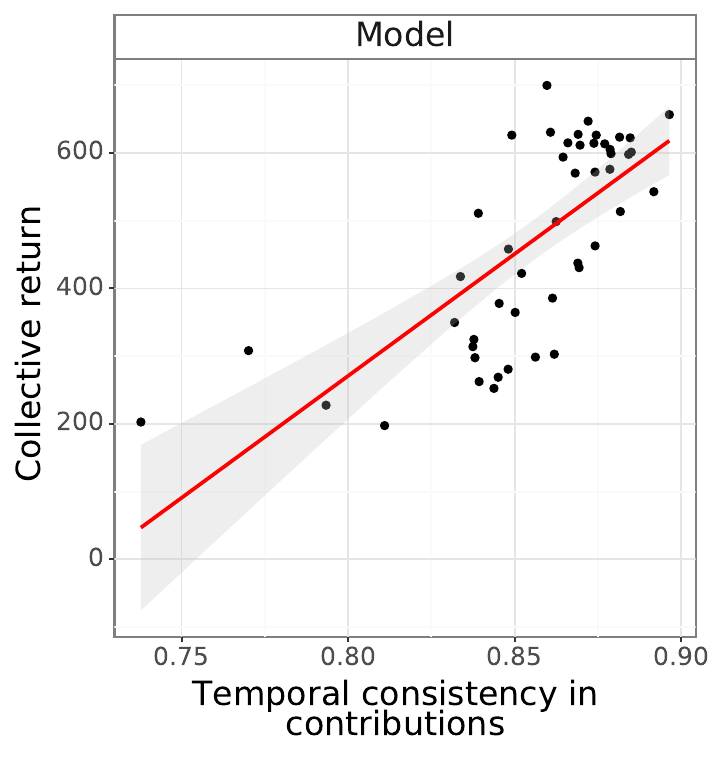}
    \caption{Figure \ref*{fig:agent_consistency}: In the model, the temporal consistency of group contributions was significantly and positively associated with group performance. Error band represents 95\% confidence interval.}
    \label{fig:agent_consistency}
\end{figure}

{With this regression, the model predicted a significant relationship between contribution consistency and collective return, $\beta = 3601.7$, 95\% CI $[2569.5, 4633.9]$, $p < 0.0001$ (Figure \ref{fig:agent_consistency}). Contribution consistency positively related to collective return, such that groups that provided the public good with greater consistency over time tended to achieve higher scores.}

\subsection{Human Behavioral Experiment}

\begin{figure}[!t]
    \centering
    \includegraphics[width=8cm]{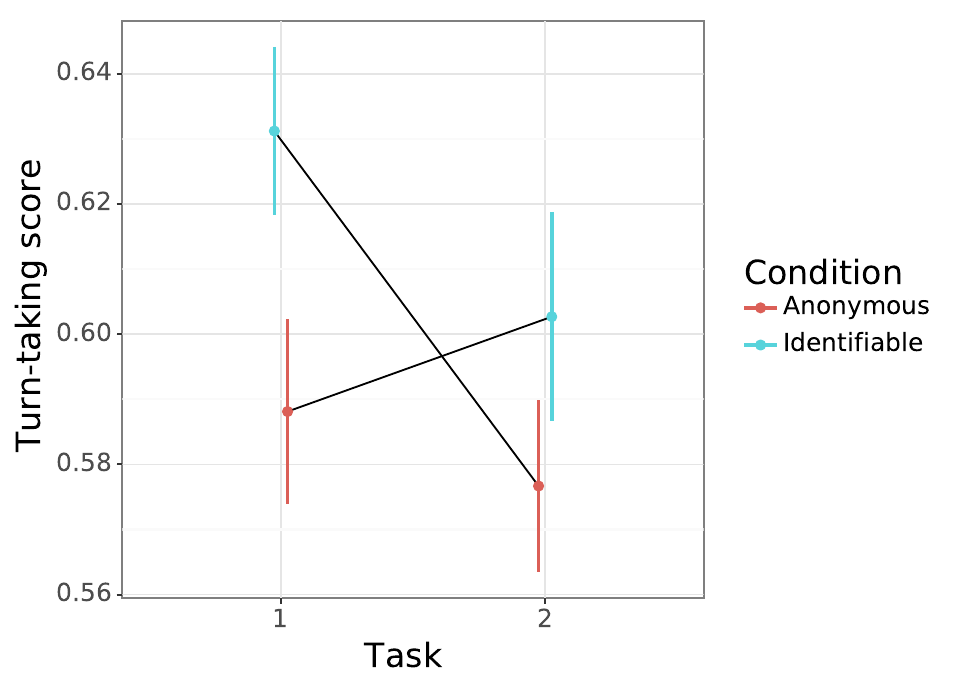}
    \caption{Figure \ref*{fig:anova_turn_taking}: We use a two-way, repeated-measures ANOVA to evaluate the effects of condition and task number on group turn taking. There was a significant main effect of condition on group turn taking. The main effect of task number was also significant, while the interaction effect between condition and task number was not significant. Error bars reflect 95\% confidence intervals.}
    \label{fig:anova_turn_taking}
\end{figure}

We conduct a two-way ANOVA to assess the effect of the intrinsic motivation for reputation on group turn taking (Figure \ref{fig:anova_turn_taking}). As before, we highlight the main effect of condition in the main text to facilitate comparison with the model results, and here expand on the other terms of the two-way ANOVA. There was a significant main effect of condition on turn taking, $F(1,310) = 29.4$, $p < 0.0001$. The main effect of task number was also significant, $F(1,310) = 9.8$, $p = 0.0019$. The interaction effect was non-significant, $F(1,22) = 0.2$, $p = 0.65$. Groups were significantly more reliant on a turn-taking rotation scheme in the identifiable condition (with an average turn-taking score of 0.62) than in the anonymous condition (with an average score of 0.58).

{In the main text, we analyze the association between turn taking and collective return with a linear regression, averaging observations by group:}

\begin{equation}
   \textrm{Collective Return} = \beta_0 + \beta_1 \cdot \textrm{Turn-Taking Score} + \epsilon \, .
\end{equation}

{Among the human groups, there was a positive relationship between turn taking and collective return, $\beta = 3784.6$, 95\% CI $[1616.8, 5950.4]$, $p = 0.0010$. The use of a turn-taking rotation scheme was positively associated with group performance.}

\begin{figure}[!b]
    \centering
    \includegraphics[width=11cm]{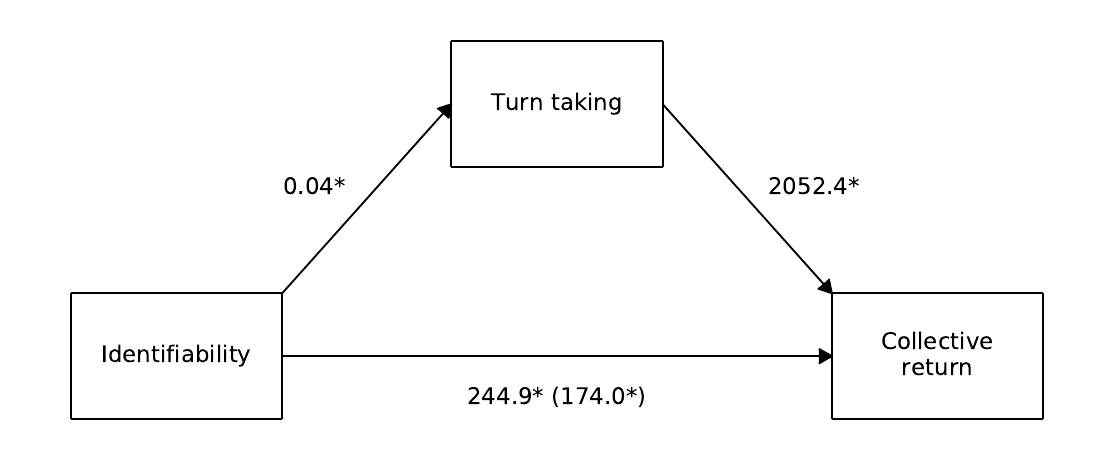}
    \caption{Figure \ref*{fig:human_turn_taking}: Mediation analysis revealed a significant indirect effect of identifiability on collective return, mediated by group turn taking. * $p < 0.05$.}
    \label{fig:human_turn_taking}
\end{figure}

{We conduct a mediation analysis to estimate the indirect effect of identifiability on collective return through group turn-taking (Figure \ref{fig:human_turn_taking}). The analysis indicates a significant and positive indirect effect of identifiability on collective return through group turn-taking, $AB = 70.9$, 95\% CI $[37.2, 112.4]$, $p < 0.0001$. In addition, the positive association between identifiability and collective return ($C = 244.9$, 95\% CI $[143.7, 338.6]$, $p < 0.0001$) is reduced after accounting for turn taking ($C' = 174.0$, 95\% CI $[72.4, 272.6]$, $p < 0.0001$).}

\begin{figure}[ht]
    \centering
    \includegraphics[width=8cm]{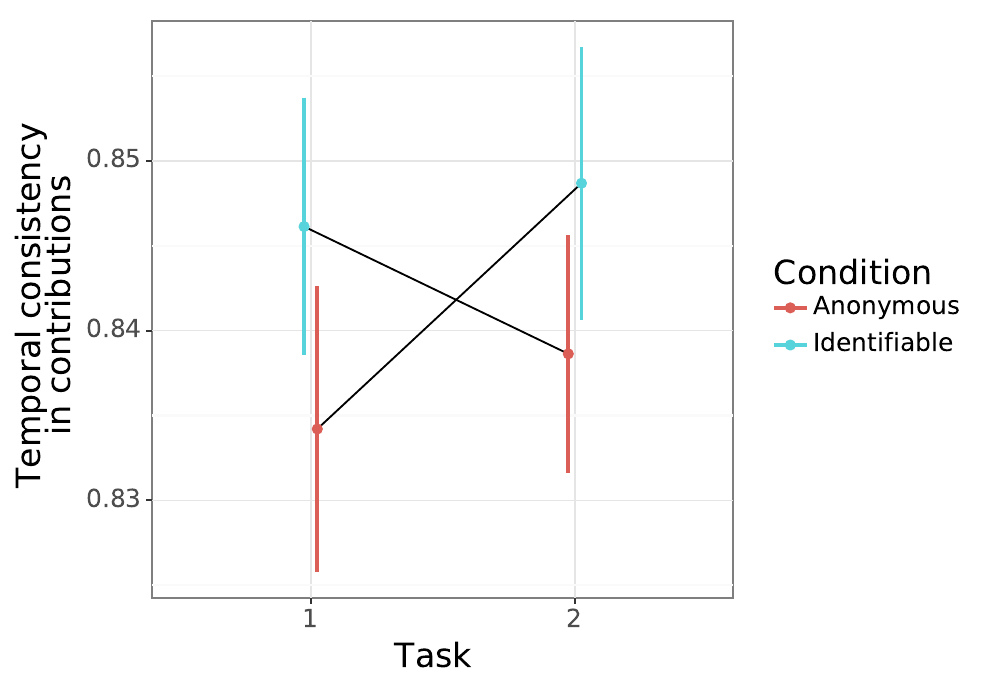}
    \caption{Figure \ref*{fig:human_anova_consistency}: We use a two-way, repeated-measures ANOVA to evaluate the effects of condition and task on temporal consistency in group contributions. There was a significant main effect of condition on temporal consistency. Neither the main effect of task number nor the interaction effect between condition and task number were significant. Error bars reflect 95\% confidence intervals.}
    \label{fig:human_anova_consistency}
\end{figure}

{To further test our findings, we replicate these analyses with the temporal consistency measure. We conduct a two-way, repeated-measures ANOVA to assess the effect of the intrinsic motivation for reputation on group contribution consistency over time (Figure \ref{fig:human_anova_consistency}). There was a significant main effect of condition on temporal consistency, $F(1,310) = 9.8$, $p = 0.0019$. The main effect of task number was not significant, $F(1,310) = 1.0$, $p = 0.32$. The interaction effect was non-significant, $F(1,22) = 0.0$, $p = 0.95$. Groups acted with greater consistency in the identifiable condition (with an average consistency score of 0.85) than in the anonymous condition (with an average score of 0.84).}

{We next evaluate the relationship between temporal consistency and group performance with a linear regression, averaging observations by group:}

\begin{equation}
   \textrm{Collective Return} = \beta_0 + \beta_1 \cdot \textrm{Contribution Consistency} + \epsilon \, .
\end{equation}

\begin{figure}[ht]
    \centering
    \includegraphics[width=6cm]{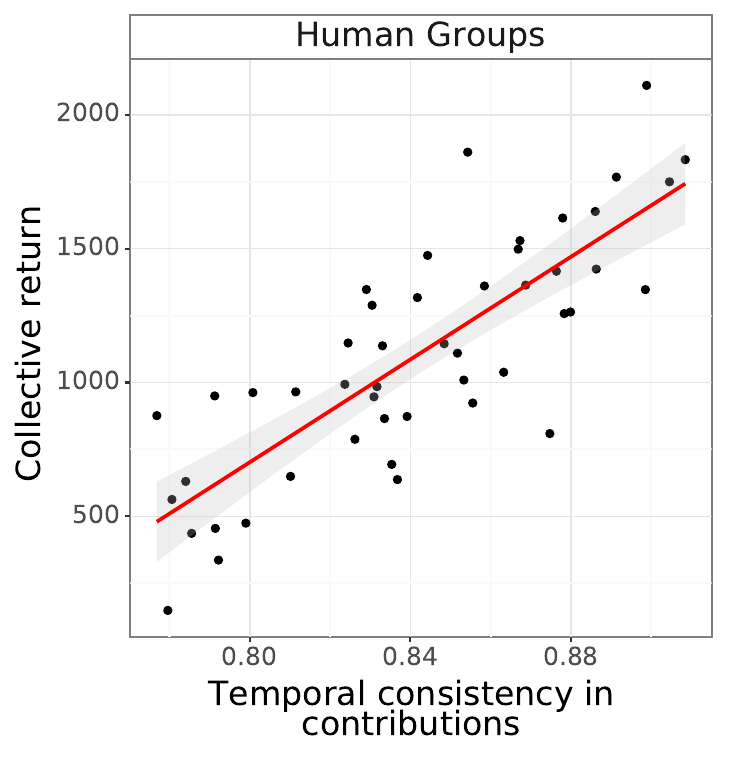}
    \caption{Figure \ref*{fig:human_consistency}: Among human groups, the temporal consistency of group contributions was significantly and positively associated with group performance. Error band represents 95\% confidence interval.}
    \label{fig:human_consistency}
\end{figure}

{Contribution consistency was significantly and positively related to collective return, $\beta = 9596.6$, 95\% CI $[7537.1, 11656.0]$, $p < 0.0001$ (Figure \ref{fig:human_consistency}). Contribution consistency positively correlated with group performance, such that groups that provided the public good with greater consistency over time tended to achieve higher scores.}

\end{document}